\documentclass{article}
\usepackage[margin=1in]{geometry}
\usepackage[super]{natbib}
\usepackage{cite}
\usepackage[pdftex]{graphicx}
\DeclareGraphicsExtensions{.pdf,.png}
\usepackage{array}
\usepackage{stfloats}
\usepackage{setspace}
\usepackage{amsmath}
\usepackage{url}
\usepackage{booktabs}

\begin{document}
\pagenumbering{gobble}
\newcommand{\dint}{\mathrm{d}}

\begin{centering}
\clearpage\thispagestyle{empty}
\textbf{\Large{Estimating the Resilience of Non-Stationary Systems}}

\vspace{2cm}

Taylor Smith$^{1}$*, Andreas Morr$^{2,3}$, Christof Sch\"{o}tz$^{3,4}$, Niklas Boers$^{3,4}$\\
$^{1}$Institute of Geosciences, Universit\"{a}t Potsdam, Potsdam, Germany\\
$^{2}$Department of Mathematics, School of Computation, Information and Technology, Technical University of Munich, Munich, Germany  \\
$^{3}$Potsdam Institute for Climate Impact Research, Potsdam, Germany\\
$^{4}$Munich Climate Center and Earth System Modelling Group, Department of Aerospace and Geodesy, TUM School of Engineering and Design, Technical University of Munich, Munich, Germany  \\

\end{centering}

\vspace{15cm}

\noindent
Corresponding author: \\
Taylor Smith \\
Email: tasmith@uni-potsdam.de

\clearpage
\doublespacing
\pagenumbering{arabic}

\begin{abstract}

A wide body of work has applied the concept of critical slowing down to estimate the stability of different Earth system components. Most of them -- such as global vegetation -- are inherently non-stationary, for example due to strong seasonal forcing, which complicates the estimation of their resilience to external perturbations. Here, we introduce a new method to account for non-stationarity in estimating resilience for diverse synthetic and real-world data sets via a regression-based formulation of the Langevin Equation. Our method does not require extensive data pre-processing, is robust to gaps in the data record, and does not require regular time sampling. We further show that our method can incorporate time-varying data uncertainties, recover uncertainty bounds in stability estimates, and can be natively extended to examine spatial systems. Our method is a drop-in replacement for widely-used autocorrelation-based resilience estimates, and can be widely applied across Earth system components.  

\end{abstract}

\clearpage
\newpage

\section*{Introduction}

There has been substantial recent research focused on understanding the stability and resilience of different Earth system components \citep{Lenton2008,boers2022_erl,lenton2024,forzieri2022,boers2021,boers2021b,boers2025}; much of this work is based on the concept of `Critical Slowing Down' (CSD), which uses slowing system dynamics when approaching critical transition points to provide a warning of oncoming state changes \citep{Carpenter2006,Dakos2008,Scheffer2009,boers2022_erl}. Lag-1 autocorrelation (AC1) and variance are the most commonly used early warning indicators, though there exist several other approaches such as spatial indicators \citep{dakos2010}, empirical and regression-based estimates \citep{smith2022,boers2021b}, eigenvalue-based approaches \citep{smith2025,grziwotz2023, morr2024b}, spectral properties \citep{bury2020,morr2024}, and deep learning \citep{bury2021,huang2024}.

The majority of CSD metrics rely on (1) evenly-sampled time series and (2) the assumption of data stationarity (i.e., that it is trend- and seasonality-free). While many instrument records are inherently evenly sampled, this is not always the case when considering, for example, paleoclimate proxy records or using composite products based on multiple data sets. Data stationarity is often accomplished by pre-processing seasonal or otherwise non-stationary data (e.g., paleoclimate records with Milankovich cycles \citep{boers2022_erl}) using statistical deseasoning and detrending techniques \citep{smith2023b, STL}. The reliance on stationary data is due to the core assumption that a system can be linearized around a stable fixed point, and hence can define an Ornstein-Uhlenbeck process \citep{Djikstra2013} with a damping rate controlling how fast a given external perturbation decays back towards the current equilibrium \citep{boers2025}. This framing motivates the use of, e.g., AC1 and variance as early-warning signals of oncoming critical transitions, and has been widely applied across disciplines \citep{Scheffer2009,lenton2024,boers2025,grziwotz2023}.

Beyond such equilibrium assumptions, existing CSD frameworks require error-prone pre-processing \citep{smith2023b,rietkerk2025}; significant issues with the calculation of AC1 and variance on real-world and gappy data have also been found in recent work \citep{Liu2026}. Furthermore, processes independent of stability loss can also drive changes in AC1 and variance, for example changes in sensor signal-to-noise ratios \citep{smith2022b,rietkerk2025}. Current methods also struggle to integrate data uncertainty into stability estimates \citep{benyami2023,benyami2024}; this is a critical limitation for many data sets -- such as paleoclimate proxies -- which have inherent and time-varying uncertainties in their measurements as well as in the corresponding age estimates \citep{Boers2017} that are often not captured using current approaches. The propagation of deseasoning and detrending errors into stability estimates is also not common practice, despite clear evidence that the choice of pre-processing strategy influences these estimates \citep{smith2023b,smith2025}. 

In this work, we introduce a regression-based approach to estimating the resilience (quantified in terms of the recovery rate $\lambda$) of a given system using the Langevin equation that can natively handle irregularly sampled data, non-stationary data, data with time-varying uncertainties, and spatially extended systems. We first motivate our approach within CSD theory, and illustrate its advantages using a synthetic time series model. We further apply our direct recovery rate estimation method to real-world data that has, until now, required several pre-processing steps, including vegetation dynamics, paleoclimate proxies, and glacier surges, and show how it can add additional context, such as uncertainty estimates, to the widely-used stability indicator $\lambda$. 

\section*{Theoretical Framing}

A general framing of the stability of a dynamical system can be motivated by assuming that a system state $x(t)$ is a stochastic process characterized by a time-dependent attractor $\mu(t)$; that attractor $\mu(t)$ can be fixed, drifting, or oscillating (i.e., a limit cycle). The dynamics of that system can be described using the Langevin equation: 

\begin{equation}
\frac{dx}{dt} = f(x,t) + \sigma(t) \eta(t)
\label{eq1}
\end{equation}

\noindent where the deterministic function $f(x,t)$ yields the time-dependent attractor and describes the restoring force against perturbations and $\eta(t)$ are Gaussian white noise disturbances with magnitude $\sigma(t)$ (i.e., noise intensity). As $\mu(t)$ captures the system's deterministic trajectory, evaluating $f(x,t)$ exactly on the attractor yields the attractor's trajectory gradient, here termed $\frac{d\mu}{dt}$. If we assume that the system is near the attractor $\mu(t)$, the system's fast and noisy dynamics are largely governed by the restoring forces encoded by $f(x,t)$ around $\mu(t)$. These forces can be described as approximately linear at each point in time, yielding:

\begin{equation}
\frac{dx}{dt} = \frac{d\mu}{dt} - \lambda(t) \big( x(t) - \mu(t) \big) + \sigma(t) \eta(t)
\label{eq2}
\end{equation}

\noindent where $\lambda(t)$ is the restoring rate (i.e., gradient of the restoring force), or how strongly perturbations decay towards the attractor. It is this linear restoring force that is typically regarded as a measure of system resilience, with a decrease towards zero signaling a loss of resilience. Equation \ref{eq2} is the basis for most common techniques of estimating $\lambda$, as removing the suspected position of the attractor from time series data yields a process that should restore towards equilibrium. In such a process, $\lambda$ is directly accessible via proxy statistics such as lag-1 autocorrelation (AC1) and variance. However, the attractor location $\mu(t)$ is typically not known and its movement $\frac{d\mu}{dt}$ is often disregarded, meaning that any attempt to remove non-stationary components by means of detrending or deseasoning can bias $\lambda$ estimates in complex ways.

Here, we propose taking the functional form of the attractor as a starting point. In many natural systems (such as seasonally forced systems), the attractor's movement is well-constrained and can be functionally modeled. The instantaneous change $f(x,t)$ in the system state can thus be approximately decomposed into a part that depends on $x$ explicitly and linearly, and a part that arises purely from external forcing and is thus a function of time $t$.

To analyze changes in $\lambda$ through time, we construct a set of moving (temporal) windows. Within each moving window, we assume our system dynamics $f$ to be quasi-stationary (i.e., the resilience $\lambda$ and noise strength $\sigma$ are constant). This means that while the system can be globally non-stationary (e.g., seasonal cycles or trends, including changes in resilience), we assume that any slow external changes (e.g., climate change) can be neglected over one temporal observation window, whose size needs to be chosen according to the speed of the system dynamics. We also assume that external perturbations (e.g., rainfall over a spatial domain) are homogeneous; this allows us to treat spatial variations in $x(t)$ as being caused by random noise and the internal dynamics and feedbacks responding to that noise, rather than by spatial variability in external perturbations. This is trivially the case for the one-dimensional time series, but must be accounted for when considering spatio-temporal data. 

Under the above assumptions, we can treat the restoring rate $\lambda(t)$ as approximately constant within a short temporal window. We can thus discretize Equation \ref{eq2} in time using the observations $x_1, \dots, x_n$ taken at time points $t_1, \dots, t_n$:

\begin{equation}
\frac{\Delta x_i}{\Delta t_i}:=\frac{x_{i+1}-x_i}{t_{i+1}-t_i} \approx \underbrace{-\lambda}_{\beta_1} x_i + \underbrace{\Big[ \lambda\mu(t_i) + \frac{d\mu}{dt}(t_i) \Big]}_{m(t_i)} + \epsilon_i
\label{lambda}
\end{equation}

\noindent where $\frac{\Delta x_i}{\Delta t_i}$ is the discrete rate of change of $x$ and $\epsilon_i$ is discrete stochastic noise; we group the attractor dynamics ($\mu$ and $\frac{d\mu}{dt}$) together for clarity as they both depend on time $t_i$. This formulation yields a linear regression problem where $\lambda$ is the negative of the slope of the regression ($\beta_1$), and both the position and movement of the attractor are contained within $m(t_i)$. The residual term $\epsilon_i$ captures any deviations from the estimated attractor $\mu$. A diminishing contribution of the system state to the regression fit -- i.e., a decreasing estimate of $\lambda$ over several temporal windows -- can be interpreted analogously as a decrease in system resilience. For our analysis, we assume that the strength of process noise ($\sigma \eta$) is substantially larger than any observation or measurement noise ($\delta$). As we rely on linear regression to estimate $\lambda$, any measurement noise $\delta$ in our independent variable $x$ will bias $\lambda$ downwards; if we assume that $\sigma \eta \gg \delta$, we can interpret $\lambda$ as a conservative (upper bound) estimate of stability. 

\subsection*{Simultaneous Estimation of $\lambda$ and $\mu$}

For a simple system with no expressions of seasonality or linear trends, any drift in the system collapses into the regression intercept ($\beta_0 = m(t_i)$), which contains both $\lambda$ and the attractor $\mu$. This assumes that the external forcing is negligible within the analysis window (e.g., slow climate change) and that there is no inherent non-stationarity in the system (i.e., it does not have any seasonal oscillations or trends).

Most systems on Earth are, however, not so simple -- many systems are driven by oscillating forcing, such as the daily or annual cycles in temperature and sunlight that are central to their dynamics. To capture such an oscillating attractor, we need to correctly pose the linear regression problem (Equation \ref{lambda}); for the case where $\mu$ contains a seasonal component with a known period (i.e., one year), we can approximate it with a Fourier series of $k$ harmonic terms:

\begin{equation}
\frac{\Delta x_i}{\Delta t_i} = \underbrace{-\lambda}_{\beta_1} x_i + \underbrace{\sum_{k=1}^{n} \Big[ A_k \sin(k\omega t_i) + B_k \cos(k\omega t_i)\Big] + \beta_0}_{m(t_i)} + \epsilon_i
\label{harmonic}
\end{equation}

\noindent where $m(t_i)$ is a time-varying intercept modeled using $A_k$ and $B_k$ as the harmonic fitting terms for frequency $\omega$ that constitute the linear regression factors beyond the linear restoring rate $\lambda$. As the derivative of a harmonic series yields another harmonic series, both $\lambda\mu(t_i)$ and $\frac{d\mu}{dt}(t_i)$ are absorbed into the coefficients $A_k$ and $B_k$; $\beta_0$ captures any constant offset in the attractor position. We choose $n\ge2$ to allow  for asymmetric seasonality (e.g., steeper greening than browning phases in vegetation) in complex real-world data. We note that Equation \ref{lambda} can also be reformulated to incorporate simpler (e.g., linear drift) or more complex components (e.g., Milankovich cycles in paleoclimate records) depending on the system under study and its assumed attractor shape. We incorporate the assumed attractor shape into our regression via a design matrix; this approach allows for flexibly modeling diverse systems, as well as combining multiple assumed forcing regimes (e.g., both seasonal and linear forcings on the system). 

For any chosen attractor model, we can then capture $\lambda$ via regression; in the simplest case, an ordinary least squares regression can be used. For systems with known uncertainty (e.g., with independent error estimates for each sample), we can also use a weighted regression to estimate $\lambda$. In principle, other regression models can also be substituted; here we use an outlier-minimizing robust regression (Methods) which is flexible to both small (e.g., one-dimensional time series) and large (extensive spatio-temporal grids) data.

\subsection*{Relationship to Widely-Used CSD Indicators}

Several proxy statistics for estimating the restoring rate $\lambda$ have been proposed in the literature, the most commonly used being AC1 and variance. Both of these metrics, however, can be polluted by signals which are unrelated to any changes in internal dynamics \citep{smith2022b,benyami2023,rietkerk2025}. For an Ornstein-Uhlenbeck process driven by white noise $\sigma$, $\lambda$ estimated from variance (here termed $\lambda_\mathrm{Var}$) is derived from both stability and driving noise strength ($\text{Var}(x) = \frac{\sigma^2}{2\lambda \Delta t}$); a change in variance can be equally caused by a loss of resilience ($\lambda$ moving towards zero) or increased driving noise (increasing $\sigma$). The widely-used autocorrelation-based $\lambda$ formulation (here termed $\lambda_{\mathrm{AC1}}$) is estimated as $\mathrm{AC1} = e^{\lambda\Delta t}$; both metrics require that the system fluctuates around a constant mean (i.e., it does not have seasonal oscillations that can bias variance and autocorrelation). Further, the system must be continuously sampled at a fixed interval $\Delta t$ without gaps, as $\lambda_{\mathrm{AC1}}$ relies on sequential pairs of fixed $\Delta t$ representing a constant lag and $\lambda_{\mathrm{Var}}$ is influenced by both changes in data density (i.e., the sample variance can be biased by more summer than winter measurements) and data spacing (i.e., the amount of random noise between measurements scales with $\Delta t$, making $\lambda_{\mathrm{Var}}$ impossible to cleanly invert for changing $\Delta t$). In principle, both of these problems can be solved via data pre-processing schemes, although interpolation (either to fill gaps or to fit an even time sampling) will by construction increase autocorrelation in a time series, and the choice of deseasoning procedure can introduce complex biases in both $\lambda_{\mathrm{AC1}}$ and $\lambda_{\mathrm{Var}}$ \citep{smith2023b}. 

Our regression-based estimate of $\lambda$ (Equation \ref{lambda}), however, does not require deseasoning and detrending to stationarity; we can rather account for, e.g., seasonality via a regression design matrix (Equation \ref{harmonic}) and solve for $\lambda$ and the (seasonal) attractor $\mu$ simultaneously. Further, by using $\frac{\Delta x_i}{\Delta t_i}$ rather than constant steps ($\Delta t_i = \Delta t$), we are not bound by fixed sampling intervals; this means that we can accommodate irregularly-sampled and gappy time series without interpolation. We note, however, that our approach does not directly yield a continuous-time estimate of $\lambda$, as is the case for $\lambda_{\mathrm{AC1}}$ and $\lambda_{\mathrm{Var}}$ which arise directly from the analytical solution to an Ornstein-Uhlenbeck process \citep{Djikstra2013}. We rather recover a discrete-time, finite-difference estimate of $\lambda$, which must be converted to continuous time before comparison with $\lambda_{\mathrm{AC1}}$ and $\lambda_{\mathrm{Var}}$, as well as to correct for changes in sampling density ($\Delta t$) through time (Methods). For all analyses, we present continuous-time estimates of $\lambda$, $\lambda_{\mathrm{AC1}}$, and $\lambda_{\mathrm{Var}}$.

\begin{figure*}[!h]
\centering
\includegraphics[width=0.75\linewidth]{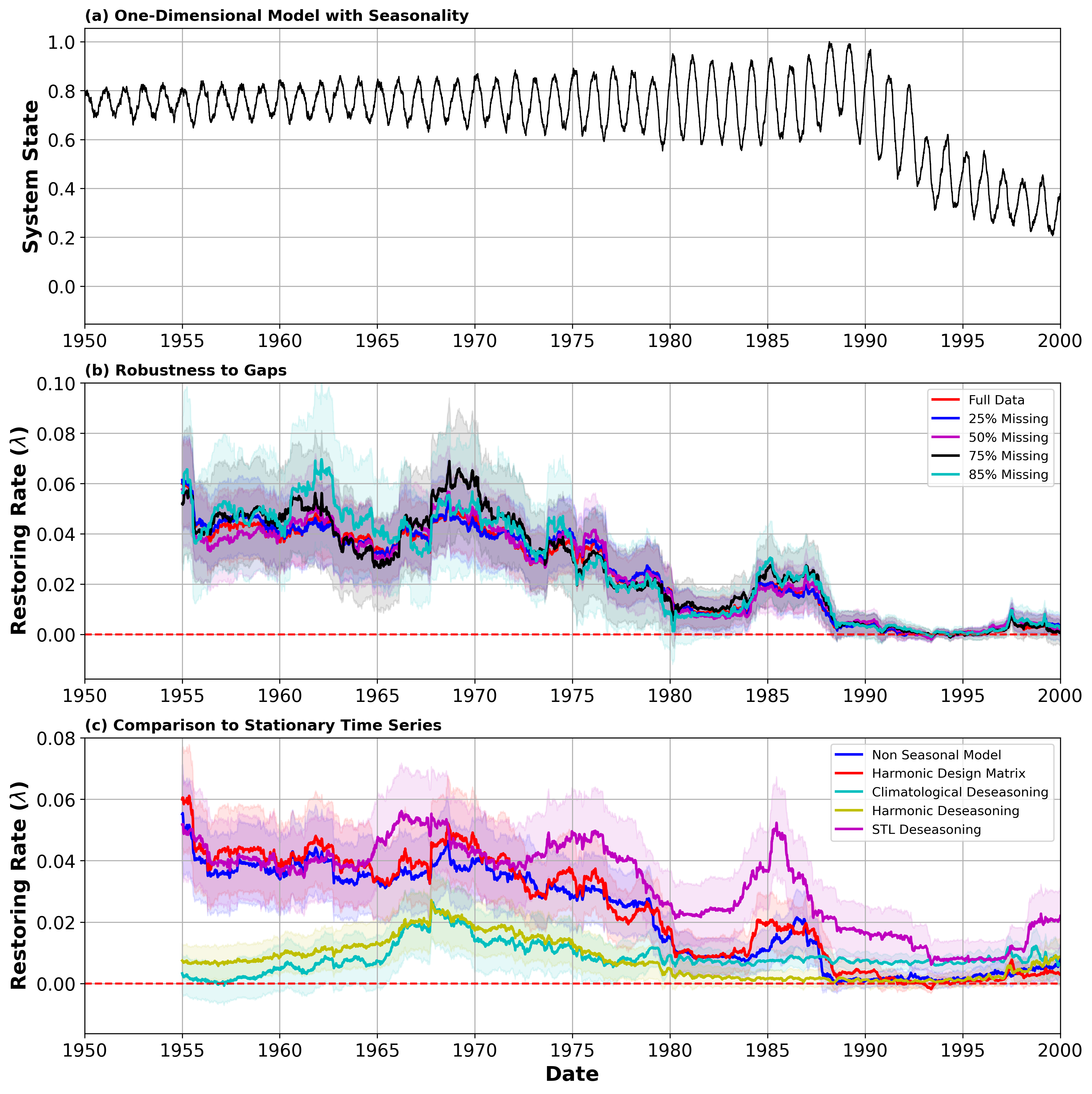}
\caption{\textbf{Stability of a simple time series model.} (A) System state. (B) Restoring rate $\lambda$ estimated using a harmonic design matrix on data sets with variable gap percentages. (C) $\lambda$ estimated on a non-seasonal control model compared to estimates on a seasonal model using a harmonic design matrix and typical deseasoning approaches. Shaded bounds (B,C) cover one standard deviation uncertainty in $\lambda$ (Methods). Note that our seasonal model using a harmonic design matrix (Methods) yields the most similar results to the non-seasonal control model.}
\label{f1}
\end{figure*}

If we consider a simple dynamical system with seasonality moving towards a state transition (Figure \ref{f1}), we can assess the robustness of the regression-based $\lambda$ to gaps and compare its performance to the common practice of first deseasoning the time series and then computing $\lambda$ on the nominally stationary residual time series. We model seasonality via harmonic terms in our regression design matrix (Equation \ref{harmonic}). 

We find that regression-based $\lambda$ is robust to gaps, even when they are extremely numerous (Figure \ref{f1}b) or long (Supplemental Figure S1). We further find that our design matrix approach successfully accounts for seasonal oscillations, and that the computed $\lambda$ on seasonal and a non-seasonal control models are very similar (Figure \ref{f1}c). This is not the case for common deseasoning methods such as Seasonal Trend Decomposition via Loess (STL) \citep{STL}, removing harmonic seasonality \citep{smith2023b}, and using the climatological mean \citep{forzieri2022}, which do not necessarily preserve the full red-noise spectrum of the underlying data and hence can substantially bias the estimation of $\lambda$ (Figure \ref{f1}c).

We can also natively extend our regression-based approach to the spatial context by collapsing the spatial dimension before regression (Methods). As in the one-dimensional case (Figure \ref{f1}), we note that our harmonic design matrix approach successfully mimics the non-seasonal control run, while common pre-processing schemes introduce significant bias into $\lambda$ estimates (Supplemental Figure S2). While the extension of our method to the spatial domain is of potentially great benefit, there are several assumptions that need to be made about the spatial heterogeneity of forcing that preclude its use in some cases (Methods).

Our regression-based approach has three further benefits beyond being robust to gaps and accounting for non-stationary attractors via a design matrix: (1) we can handle uncertainty and measurement errors robustly using a weighted regression; (2) we recover uncertainty estimates of $\lambda$ that explicitly account for errors in the assumed attractor shape; and (3) we implicitly account for uneven time sampling, removing the need for temporal interpolation. We use three illustrative examples to show how we can capture $\lambda$ in settings that previously required extensive pre-processing and data cleaning. 

\section*{Global Vegetation Resilience}

\begin{figure*}[!h]
\centering
\includegraphics[width=0.65\linewidth]{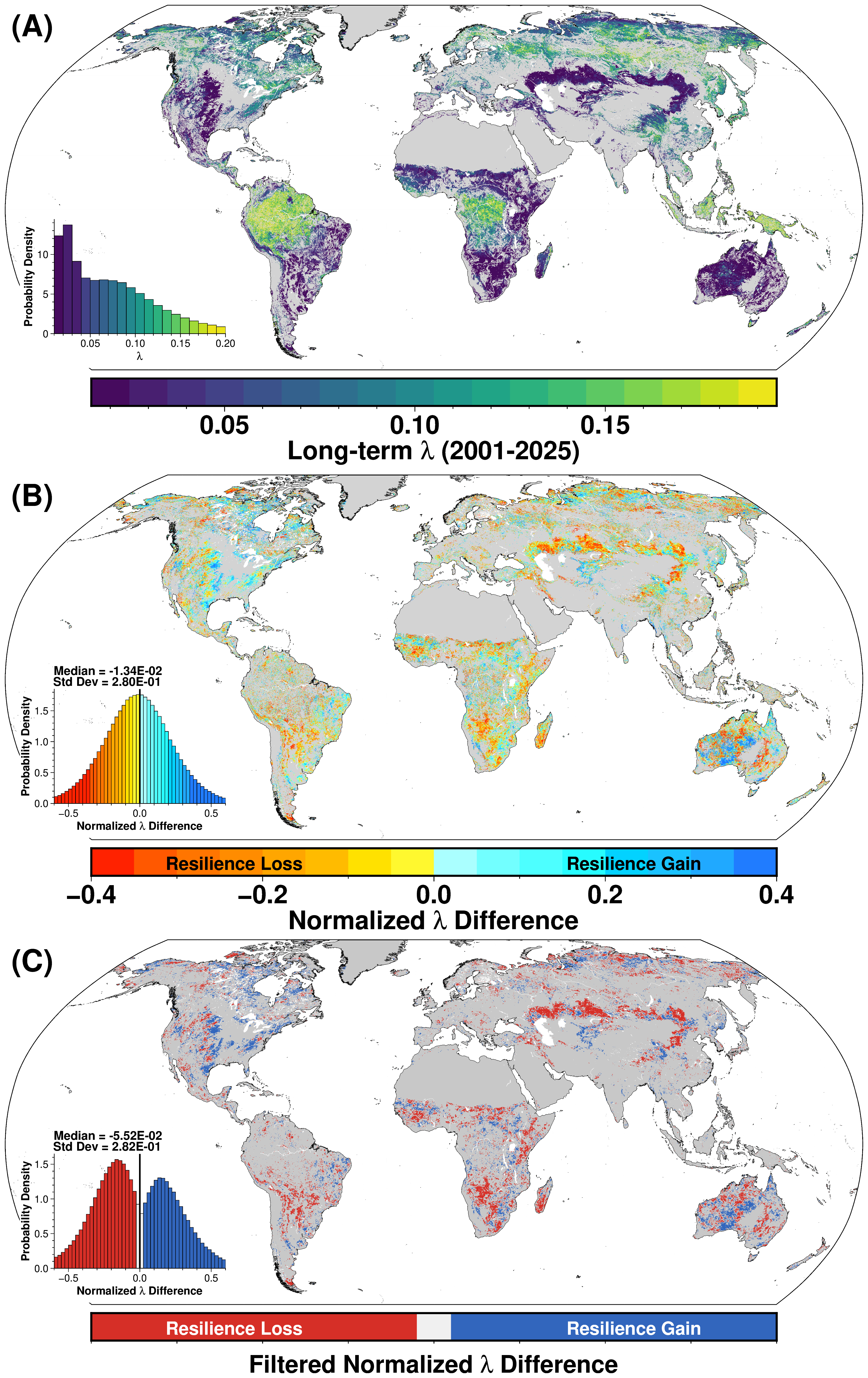}
\caption{\textbf{Global Vegetation Resilience.} (A) Estimated $\lambda$ over vegetated ecosystems (MODIS kNDVI, 2001-2025), with anthropogenic land-cover types masked based on MODIS land-cover data (Methods). (B) Normalized difference in $\lambda$ between the period 2013-2025 and 2001-2012, showing a slight tendency towards decreasing $\lambda$ globally. (C) $\lambda$ differences filtered by spatial consistency (Methods), showing fewer coherent blocks of positive and negative $\lambda$ changes globally, and a stronger tendency towards resilience loss. Insets show histograms of global values, with median and standard deviation marked for difference maps.}
\label{f2}
\end{figure*}

Several studies have examined global vegetation resilience patterns \citep{forzieri2022,smith2022,boulton2022,blaschke2024,verbesselt2016}; however, estimates of vegetation resilience change vary widely with the employed methods and datasets. There is also substantial work showing that common pre-processing strategies bias estimates of $\lambda$ \citep{smith2023b}, as well as strong evidence for gaps driving at least some of the spatial diversity in vegetation resilience trends \citep{Liu2026}. If we re-assess one commonly-used data set (kNDVI from MODIS \citep{CampsValls2021,Didan2021}), we can directly compare $\lambda$ estimates over two distinct time periods covering half of the data record each (2001-2012 and 2013-2025, Figure \ref{f2}). We use both a harmonic and a local linear term in our regression design matrix to account for both slow changes in the mean state -- such as greening or browning -- and annual seasonality. 

We confirm previous results (i.e., the broad-scale patterns of $\lambda$ and $\lambda$ decreases \citep{smith2023b}) with our method that is (1) robust to gaps (Figure \ref{f1}) and (2) handles seasonality and trends via a design matrix (Equation \ref{harmonic}) rather than statistical deseasoning techniques. We document decreasing stability in many -- but not all -- ecosystems, implying longer recovery times from perturbations, as has previously been reported \citep{forzieri2022,smith2022,smith2023b}. We further note that the spatial patterns of $\lambda$ derived using our method and via AC1 or variance are broadly similar (Supplemental Figure S3), though they are strongly influenced by the choice of deseasoning procedure. 

We note that we cannot confirm or refute large-scale changes in Amazon vegetation resilience \citep{boulton2022,blaschke2024,smith2025} that have been previously argued for based on Vegetation Optical Depth \citep{VOD,boulton2022,smith2022,blaschke2024} data. We attribute the lack of spatially coherent changes in $\lambda$ to the inherent limitations of kNDVI data in dense and cloud-covered forests where optical vegetation indices tend to saturate and lose the ability to concisely capture ecosystem dynamics.

\section*{Paleoclimate Proxy Data}

Paleoclimate proxies are a difficult data source with uneven sampling rates and diverse uncertainty estimates; previous approaches \citep[e.g.,][]{Hummel2025,Boettner2021,Boers2018} have used a range of pre-processing techniques to make the data more tractable within the critical slowing down framework. We use the raw NGRIP ice core data \citep{NGRIP,Gkinis2014}, without pre-processing or resampling, to assess changes in $\lambda$ through time; this data has been used to argue for signs of critical slowing down before Dansgaard–Oeschger (DO) events \citep{Boers2018,Mitsui2024}, which mark rapid shifts in Northern-Hemisphere climate with global imprints \citep{boers2018b}, although recent work including many other Greenland ice cores provides a more inconclusive picture regarding the presence of critical slowing down prior to the DO events \citep{Hummel2025}. In contrast to previous work, we incorporate both measurement error \citep{NGRIP} and timing uncertainty into our $\lambda$ estimates, and capture local changes in the mean state via a linear term in our regression design matrix (Methods). We further correct for uneven sampling rates in the NGRIP data to minimize biases due to variable $\Delta t_i$, especially in deeper sections of the ice core (Methods).

\begin{figure*}[!h]
\centering
\includegraphics[width=0.75\linewidth]{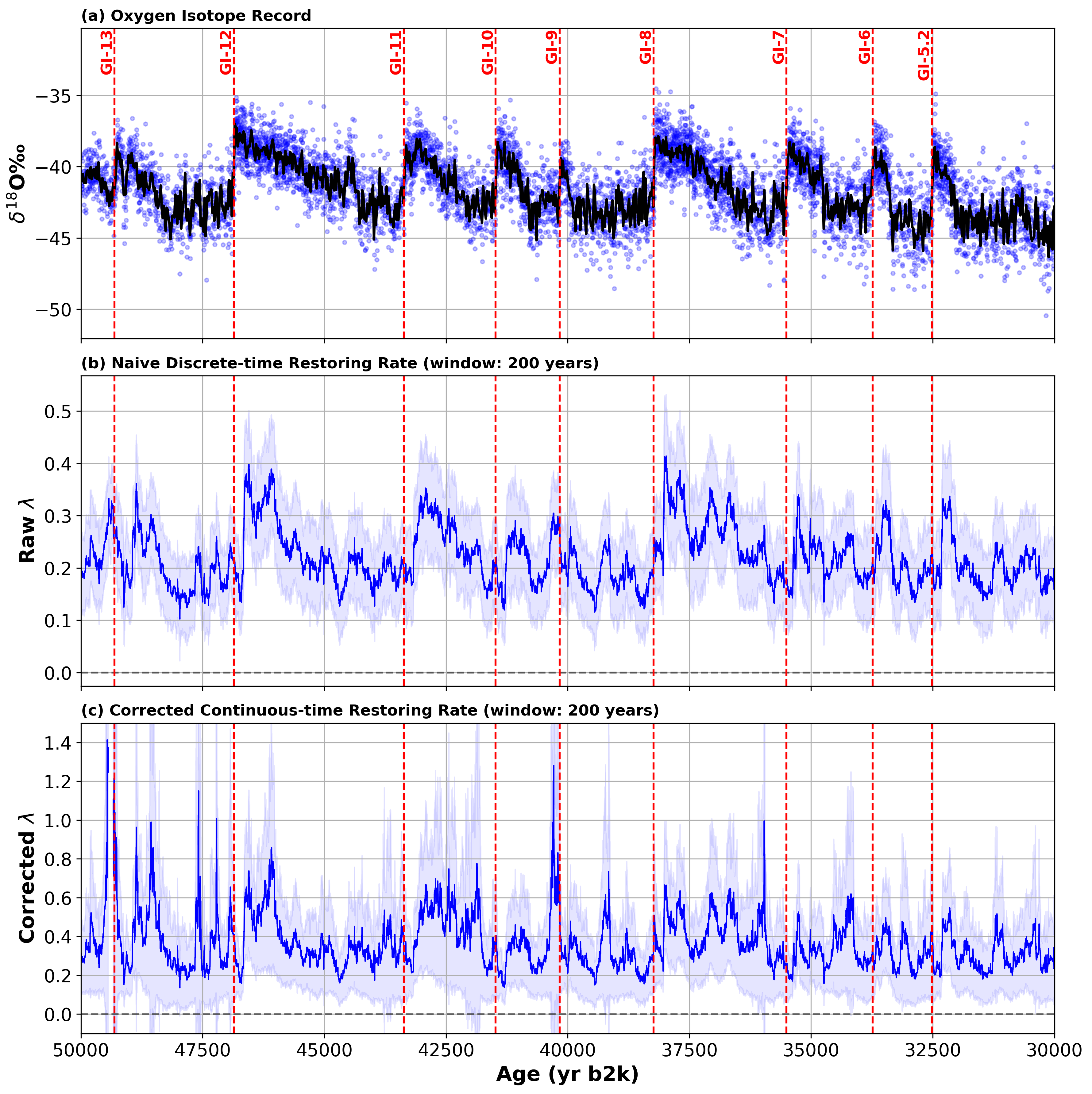}
\caption{\textbf{NGRIP Ice Core Data.} (A) $\delta^{18}\text{O}$ isotope records from NGRIP \citep{NGRIP}. Vertical lines mark DO event timing after \citep{Boers2018}. Data subset to 20,000 years for clarity. Full period of the high-resolution NGRIP data can be found in Supplemental Figure S4. (B) Changes in $\lambda$ calculated on the raw time series without pre-processing, incorporating measurement and age-model uncertainties via weighted least squares (Methods). (C) Corrected $\lambda$ accounting for changes in sampling intervals through time (Methods), showing less pronounced decreases in $\lambda$ before many DO events. Shaded bounds (B,C) cover one standard deviation uncertainty in $\lambda$ (Methods). Bias-corrected $\lambda$ estimates do not show conclusive evidence for consistent decreases in $\lambda$ before DO events.}
\label{f3}
\end{figure*}

We find inconclusive evidence for statistical precursor signals preceding DO events based on our bias-corrected $\lambda$ (Figure \ref{f3}). This agrees with previous assessments which quantified $\lambda$ in terms of increasing variance and autocorrelation \citep{boers2018b,Hummel2025}, and found statistically inconclusive evidence for early warning of DO events. We further note that $\lambda$ estimates which are not corrected for strongly variable $\Delta t_i$ (Figure \ref{f3}b) are biased in the deep sections of the NGRIP data, and provide misleading signals of critical slowing down prior to DO events due to the impacts of uneven sampling intervals (Methods). We note, however, that recent (past 2,000 years) $\lambda$ estimates are the lowest they have been in at least 50,000 years (Supplemental Figure S4), indicating a less stable Northern Hemisphere climate than in the past interglacial interval.

\section*{Glacier Surge Prediction}

Glacier surging is a well-documented phenomenon that impacts glaciers worldwide \citep{kaab2023,guillet2022}. Not all glaciers surge; whether or not there are periodic glacier surges is controlled by the interaction between downslope mass transport and basal friction forces \citep{ou2022,kaab2023,benn2019,benn2023,thogersen2019}. In this sense, a glacier can be considered a dynamical system with both steady-state downhill movement and transient surges. 

Recent advances in image cross-correlation have yielded globally resolved glacier velocity maps \citep{gardner2018,gardner2025}. These velocity maps are based on spatially and temporally overlapping pixel-offsets from multiple optical and radar sensors with variable data quality, revisit times, and spatial resolutions. Glacier velocity data is hence (1) seasonal, (2) unevenly sampled, and (3) has temporally-variable noise levels. For typical (e.g., AC1-based) approaches, substantial pre-processing is necessary; removing seasonality from highly variable glacier accumulation/ablation dynamics is also difficult \citep{smith2025}. Properly integrating the timescale of glacier velocity measurements is also not straightforward; velocity is measured over varying time spans, meaning that many measurements overlap and have different degrees of temporal smoothing. Estimating variance or autocorrelation changes is hence difficult, as there is no fixed sampling window over which to perform the analysis. 

Our regression-based approach, however, can incorporate the velocity ($v$) measurement span directly into the estimated $\frac{\Delta v_i}{\Delta t_i}$, and hence handle overlapping span-averaged measurements with the same rigor as instantaneous measurements (Methods). We note that the `naive' approach of treating velocity measurements as occurring instantaneously at their span midpoints is also tractable within our framework; the midpoint approach, however, inflates the derivative $\frac{\Delta v_i}{\Delta t_i}$ for small midpoint gaps and hence amplifies high-frequency changes in the time series. To limit biases from derivative inflation we rely on a span-aware estimate of $\lambda$ (Methods).

\begin{figure*}[!h]
\centering
\includegraphics[width=\linewidth]{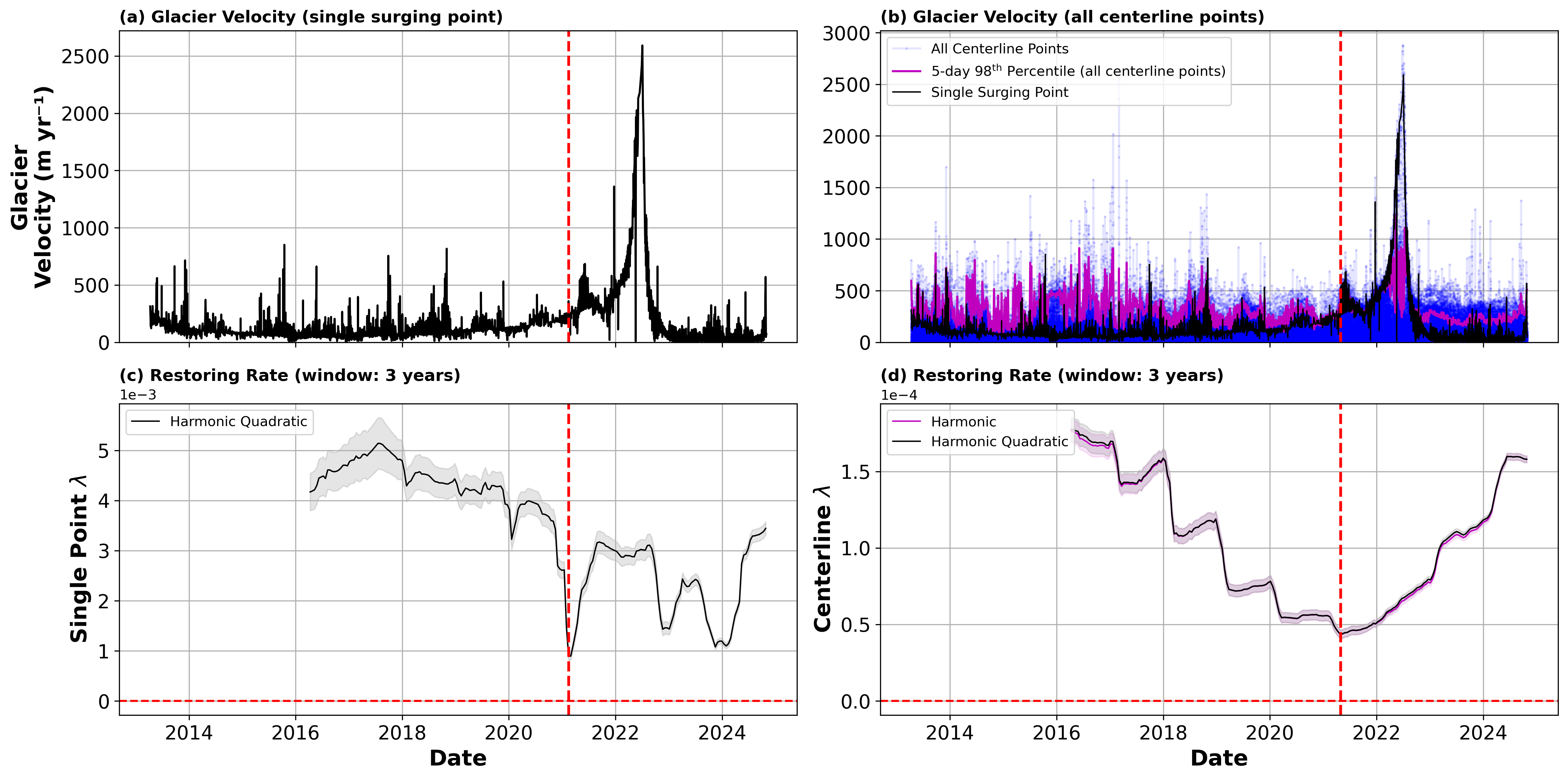}
\caption{\textbf{Stability of a surging glacier.} (A) Glacier velocity sampled at a single surging point and (B) over all glacier centerline points in 500 m steps (Methods). (C,D) Restoring rate $\lambda$, showing a rapid decline in stability before the onset of a major surge event. $\lambda$ estimated using a design matrix that includes a harmonic and quadratic term (black), as well as only a harmonic term (purple) for comparison in the 2D case. Local minima are marked with vertical dashed lines. Shaded bounds (C,D) cover one standard deviation uncertainty in $\lambda$ (Methods). Decreases in stability before the glacier surge can be seen in both the single hand-picked location and in the set of all glacier centerline points.}
\label{f4}
\end{figure*}

It is important to present a crucial detail about our glacier analysis: glacier surges inherently involve non-linear (often exponential) background acceleration. In a strict CSD sense, non-linear acceleration invalidates the basic assumption of an Ornstein-Uhlenbeck process restoring towards a stationary mean. This introduces ambiguity into the interpretation of $\lambda$ depending on how that non-linear trend is incorporated into the analysis. If we intentionally omit the non-linear term(s) from our design matrix (e.g., only using harmonic terms), the linear model does not track background non-linear acceleration. $\lambda$ dropping towards zero hence detects that drift away from the baseline mean velocity; that signal could be interpreted as a precursor of system change, but could no longer be interpreted as critical slowing down. Treating the non-linear part of glacier acceleration as strictly exponential would motivate a linear analysis on log-transformed glacier velocity data, though this approach enhances the influence of large outliers. Adding higher-order terms to our design matrix (e.g., quadratic) is less likely to bias the estimation of $\lambda$, and allows us to again interpret changes in $\lambda$ as critical slowing down, as the higher-order polynomial absorbs local non-linear trends (Figure \ref{f4}). 

There is a clear signal of stability loss ($\lambda$ approaching zero from above) before the onset of the main glacier surge phase (Figure \ref{f4}); this signal is visible in both the one- and two-dimensional cases. We attribute $\lambda$ decreases to modifications to bed friction as water builds up and glacier flow moves from a `sticky' regime to a `sliding' regime \citep{benn2019,benn2023,thogersen2019}. While a thorough examination of the precise surge initiation mechanism is beyond the scope of this study, we posit that the steep decreases in $\lambda$ before surge onset can be used to predict oncoming glacier surge activity, and $\lambda$ is linked to distinct glacier physical movement regimes. We note that in the two-dimensional case, we find qualitatively the same early-warning signal when we omit the quadratic term from our design matrix (Figure \ref{f4}d). This behavior stems from out-of-phase non-linear acceleration throughout the glacier, which minimizes the need for a quadratic term, as opposed to in the one-dimensional case where substantial biases to $\lambda$ would be expected if the quadratic term was omitted.

Recent work \citep{smith2025} has used dynamical systems theory to predict glacier surge onset; their eigenvalue-based method, however, requires complex state-space embedding parameterization, as well as resampling glacier velocity data onto a consistent temporal grid. Our regression-based approach hence provides additional benefits, particularly in that it can incorporate velocity uncertainty information directly into the estimation of $\lambda$ and that it can also integrate spatial glacier velocity data via sampling the entire glacier centerline instead of a single surging point (Figure \ref{f4}d). Treating the entire glacier as a single system, however, makes broader assumptions about the heterogeneity of external forcing (e.g., snowfall, temperature) than are required for a single time series. While the temporal pattern of inferred whole-glacier $\lambda$ is suggestive of the same pre-surge dynamics, further validation against in-situ or other surge-mapping data would be needed to confirm our findings. 

\section*{Discussion}

The estimation of $\lambda$ via linear regression provides a few key benefits over the most common approaches to estimating $\lambda$ via AC1 or variance, especially with real-world and noisy data. We can model arbitrary attractors directly in our $\lambda$ estimate via a design matrix, which makes it possible to concisely account for non-stationarity (e.g., seasonality, Figures \ref{f1}, \ref{f2}, \ref{f4}). We hence do not need to pre-process our data to remove seasonality, limiting the introduction of errors and spurious signals from the choice of deseasoning procedure. As we model both the attractor $\mu$ and $\lambda$ via a single regression, we can compute one uncertainty estimate that accounts for uncertainty in both the attractor $\mu$ (e.g., due to complex seasonal cycles) and $\lambda$. This means that we recover a direct estimate of the total uncertainty in $\lambda$, which has so far been difficult with previous methods. Furthermore, by framing our analysis around $\frac{\Delta x_i}{\Delta t_i}$, we natively account for gaps and hence do not require complete or evenly-sampled time series. This is a critical benefit for some systems, such as the vegetation (Figure \ref{f2}), paleoclimate (Figure \ref{f3}), and glacier (Figure \ref{f4}) data shown here. 

Finally, by posing our estimation of $\lambda$ as a regression problem, we inherit a very wide body of methodologies for dealing with noisy and complex data. For example, our method can directly integrate time-varying weights into our estimates of $\lambda$ by relying on weighted least squares (Figures \ref{f2}-\ref{f4}), and we can minimize the influence of outliers on our $\lambda$ estimates by robust regression approaches (Methods). We can also recover error bounds via the regression residuals -- as well as quantify the unexplained variance of $\lambda$ -- which can serve as a proxy for the external noise forcing on the system. Finally, it is possible to expand our method to the spatial domain -- with caveats -- by collapsing the spatial dimension and performing the same regression proposed in Equation \ref{lambda} over a much denser view of a given system. It would also, in principle, be possible to extend our method to include multi-variate systems, though the interpretation of a recovered multi-variate $\lambda$ would be more difficult.

Over small spatial areas with homogeneous forcing (e.g., the same weather, similar vegetation mixes), using a small spatial field instead of a single time series can provide a more data-rich view of the system, and thus a more robust regression-based estimate of $\lambda$. It is not, however, well-suited to large and complex systems, particularly those where there are substantial forcing gradients (e.g., spatially varying rainfall patterns), since it is not easy to disentangle changes in the system response to forcing ($\lambda$) from spatial variability in the forcing itself. Further research is needed to develop spatially-extended regression-based $\lambda$ estimates and rigorously compare them to established spatial early-warning signals \citep{dakos2010,smith2025}.

Our regression-based $\lambda$ estimate provides a robust means of estimating the stability of diverse systems, without key limitations that require substantial pre-processing of many data sets. We further provide a straightforward means by which to capture and account for uncertainty in real-world measurements. Our method can function as a drop-in replacement for the wide body of literature relying on autocorrelation-based stability estimates, and provides a means of simplifying and minimizing data pre-processing to increase the diversity of systems that can be assessed within a critical slowing down framework.

\section*{Methods}

\subsection*{Robust $\lambda$ Estimation via Regression}

To minimize biases induced by outliers in our regression-based $\lambda$ estimate (Equation \ref{lambda}) -- particularly in very dense and noisy time series (e.g., spatial data, paleoclimate records) -- we rely on a robust regression optimized using Iteratively Reweighted Least Squares. We downweight outliers in our data using Huber's T, which linearly suppresses the influence of large residuals after an initial least squares fit. In cases where we have access to time-explicit uncertainties (i.e., Figure \ref{f3}, \ref{f4}), we incorporate them by scaling both the design matrix and the dependent variable by the square root of the weights. We also tested a naive Ordinary Least Squares and Generalized Least Squares which accounts for autocorrelation \citep{boers2021b}, finding that they produced similar results for our synthetic data (Supplemental Figure S5). As the regression estimate of $\lambda$ is auto-regressive, there will be a small sample bias of order $O(1/n)$ \citep{Shaman1988}; for the relatively large $n$ we use here, we do not explicitly incorporate this bias. 

We capture the uncertainty in our estimated $\lambda$ using the standard error ($SE_\lambda$) of the regression slope. We construct a 95\% confidence interval around $\lambda$ ($\lambda \pm 1.96 \cdot SE_{\lambda}$); if at any point that confidence interval crosses zero, we can no longer exclude the possibility that the system has become unstable. The standard error we report (confidence bounds on Figures \ref{f1}, \ref{f3}, and \ref{f4} $\lambda$ estimates) is expanded to account for serial correlation in the time series data (reduced degrees of freedom) using the temporal autocorrelation of the regression residuals. Code to reproduce our $\lambda$ estimation procedure can be found on Zenodo \citep{smith2026_code}. 

\subsection*{Error Propagation and Uncertainty}

The most common approach to estimating $\lambda$ on non-stationary data is to first pre-process the data to remove, e.g., seasonality, and then estimate $\lambda$ on the cleaned, nominally stationary, residuals (e.g., via AC1). Our approach of using a single design matrix to capture attractor movement and $\lambda$ yields two key benefits with regards to uncertainty quantification: covariance partitioning and straightforward error propagation. 

An underlying and underappreciated assumption of pre-processing strategies is that the two signals of interest (e.g., seasonality and system dynamics) are orthogonal; that is, they can be perfectly decomposed. This could be the case for synthetic systems with, e.g., additive seasonality, but is unlikely to occur in real-world systems. It is often the case that the initial drift towards instability is attributed to the seasonal attractor, and the recovered deseasoned residuals are flatter than they should be. By instead performing a single regression with a robust design matrix, the regression coefficients for $\lambda$ and $\mu$ (e.g., Equation \ref{harmonic}) are both tuned to minimize residual variance simultaneously. This means that if, e.g., glacier acceleration is not strictly periodic, it is correctly partitioned into the $\lambda$ term rather than as part of normal seasonal glacier acceleration. We note that our design-matrix approach would be strictly equivalent to performing two linear regressions -- one to remove the attractor and one to estimate $\lambda$ -- under the condition that ordinary least squares is used for both steps (i.e., the Frisch-Waugh-Lovell theorem \citep{Frisch1933}). This equivalence is violated by most commonly used pre-processing steps such as removing climate anomalies (e.g., long-term monthly means) or using more complex deseasoning procedures (e.g., STL).

Estimates of $\lambda$ are generally taken at face value -- there are no uncertainty bounds when $\lambda$ is computed from AC1 or variance. Furthermore, errors due to the deseasoning and detrending process (e.g., partitioning too much signal into the seasonal term) are not propagated forward into estimates of $\lambda_{\mathrm{AC1}}$ or $\lambda_{\mathrm{Var}}$. In our approach, we derive the standard error of $\lambda$ directly from the covariance matrix of the regression, which includes the covariance between all of the components of the design matrix (e.g., seasonality) and our estimate of $\lambda$. Hence, if $\mu$ is poorly constrained in our regression (e.g., due to large gaps), we propagate that uncertainty into our derived standard error of $\lambda$; we thus account for errors in both our modeled attractor $\mu$ and $\lambda$ itself. In practice, this means that poor data quality (gaps, large measurement errors) will inflate the standard error, yielding wider uncertainty bounds on $\lambda$. 

\subsection*{Continuous-time Estimation of $\lambda$}

A common basis for estimating $\lambda$ is to treat a system as an Ornstein-Uhlenbeck process:

\begin{equation}
\frac{dx}{dt} = -\lambda_c (x(t) - \mu) + \sigma_c \eta(t)
\end{equation}

\noindent where $\mu$ is a stable equilibrium state, $\sigma_c$ is continuous noise strength, and $\eta(t)$ is a Gaussian white noise. In this formulation, $\lambda_c$ is the true continuous-time restoring rate against perturbations to the system. Any data-driven estimate of $\lambda_c$ will -- by necessity -- be discretized, as we recover $\lambda_c$ from a time series sampled against some continuous process; we term this discrete-time estimate as $\lambda_d$. In order to convert $\lambda_d$ to $\lambda_c$ consistently for multiple means of estimating $\lambda$ (via regression, AC1, or variance), we use an analytical conversion incorporating the data sampling interval. We define the window-averaged time sampling rate as $\langle \Delta t \rangle$, and use it to convert $\lambda_d$, $\lambda_{d,\mathrm{AC1}}$, and $\lambda_{d,\mathrm{Var}}$ to their continuous-time equivalents. 

For our regression-based $\lambda$, equating the discrete-time autocorrelation to the exact analytical solution for a single increment of Equation \ref{eq2} with time step $\Delta t$ yields:

\begin{equation}
\lambda_d = \frac{1 - e^{-\lambda_c  \Delta t }}{ \Delta t }
\end{equation}

\noindent If we expand this via a Taylor series, we find a clear inherent sampling bias in the discrete-time $\lambda_d$ versus the continuous-time $\lambda_c$:

\begin{equation}
\lambda_d = \frac{1 - \left( 1 - \lambda_c \langle \Delta t \rangle + \frac{1}{2}\lambda_c^2 \langle \Delta t \rangle^2 - \mathcal{O}(\langle \Delta t \rangle^3) \right)}{\langle \Delta t \rangle}
\end{equation}

\noindent or, simplified:

\begin{equation}
\lambda_d \approx \lambda_c - \frac{1}{2}\lambda_c^2 \langle \Delta t \rangle
\end{equation}

\noindent The directly-estimated $\lambda_d$ can be strongly biased (underestimates resilience) for large $\langle \Delta t \rangle$ -- as commonly occurs in paleoclimate data (Figure \ref{f3}) -- and higher-order terms ($\mathcal{O}$) cannot be safely ignored. We thus rather invert the full analytical relationship between $\lambda_d$ and $\lambda_c$ as:

\begin{equation}
\lambda_{c} = \frac{-\ln(1 - \lambda_d \langle \Delta t \rangle)}{\langle \Delta t \rangle}
\end{equation}

\noindent yielding an estimate of $\lambda_c$ that is less sensitive to changing sampling rates through time than the directly-estimated $\lambda_d$. 

For a discretely-sampled time series, the AC1 decays exponentially as:

\begin{equation}
\mathrm{AC1} = e^{-\lambda_c \Delta t}
\label{ac1}
\end{equation}

\noindent In order to convert $\lambda_{\mathrm{AC1}}$ to continuous-time, we incorporate the mean sampling rate $\langle \Delta t \rangle$ as:

\begin{equation}
\lambda_{c,\mathrm{AC1}} = \frac{-\ln(\mathrm{AC1})}{\langle \Delta t \rangle}
\end{equation}

\noindent We note that $\lambda_{\mathrm{AC1}}$ is only strictly valid for evenly-sampled data; by correcting with $\langle \Delta t \rangle$, we can mitigate missing data biases and transform the units of $\lambda_{c,\mathrm{AC1}}$ to match those of $\lambda_c$, which are in physical-time units (e.g., restoring rate per year).  

The conversion for $\lambda_{c,\mathrm{Var}}$ is slightly more complex in that it incorporates both variance in the state ($\text{Var}(x)$) and the variance of the driving noise over each discrete time step. For a continuous-time Ornstein-Uhlenbeck process, the variance is related to $\lambda$ as:

\begin{equation}
\text{Var}(x) = \frac{\sigma_c^2}{2\lambda_c}
\end{equation}

\noindent For a discrete time sampling $\langle \Delta t \rangle$, the system behaves as an AR(1) process and the discrete noise variance ($\sigma_d^2$) -- obtained as residuals from an AR(1) regression (used to calculate AC1) -- is an approximation of the continuous noise scaled by the time step ($\sigma_d^2 \approx \sigma_c^2 \langle \Delta t \rangle$). We can use that conversion, as well as the exponential decay of the AC1 (Equation \ref{ac1}), to reformulate the discrete-time relationship between variance and $\lambda$ in terms of continuous variables:

\begin{equation}
\text{Var}(x) = \frac{\sigma_d^2}{1 - \mathrm{AC1}^{2}} = \frac{\sigma_c^2 \langle \Delta t \rangle}{1 - e^{-2\lambda_c \langle \Delta t \rangle}}
\end{equation}

\noindent We can then solve for $\lambda_{c,\mathrm{Var}}$ as:

\begin{equation}
\lambda_{c,\mathrm{Var}} = \frac{-\ln \left( 1 - \frac{\sigma_d^2}{\text{Var}(x)} \right)}{2 \langle \Delta t \rangle}
\end{equation}

\noindent We emphasize that as with $\lambda_{\mathrm{AC1}}$, $\lambda_{\mathrm{Var}}$ is only strictly valid for evenly-sampled data, despite the $\langle \Delta t \rangle$ correction. $\lambda_{\mathrm{AC1}}$ relies on paired neighboring observations, and does not integrate the time between the observations; $\lambda_{\mathrm{Var}}$ can be biased by both changes in sample variance through time and the integration time for random noise between measurements. Unfortunately, integrating variable $\Delta t_i$ into each pointwise AC1 and variance estimate is practically intractable, and as the theoretical physical restoring rate $\lambda_c$ is exponential, our mean-time correction $\langle \Delta t \rangle$ is not a perfect solution. Any variability in the sampling frequency will introduce a bias into both estimators; missing data will also bias $\lambda_{\mathrm{AC1}}$ and $\lambda_{\mathrm{Var}}$ away from each other \citep{Liu2026}. The conversion from $\lambda_{d,\mathrm{AC1}}$ and $\lambda_{d,\mathrm{Var}}$ to $\lambda_{c,\mathrm{AC1}}$ and $\lambda_{c,\mathrm{Var}}$ using $\langle \Delta t \rangle$ can account for shifts in the mean sampling rate, but cannot correct for variance in the sampling rate, which biases the $\lambda_d \mapsto \lambda_c$ conversions due to Jensen's inequality.

For varying time windows $\Delta t_i$, the discrete-time least squares regression estimates $\lambda_d$ are defined as $\frac{1}{n} \sum_{i=1}^n \frac{1 - e^{-\lambda_c \Delta t_i}}{\Delta t_i}$. As the function $s \mapsto \frac{1 - e^{-\lambda_c s}}{s}$ is strictly convex for $s > 0$ and $\lambda_c > 0$, the discrete-time $\lambda_d$ estimate will always be greater than or equal to $\frac{1 - e^{-\lambda_c \langle \Delta t \rangle}}{\langle \Delta t \rangle}$. Our continuous-time conversion for the regression-based $\lambda$ estimates hence yield an upper bound; the true $\lambda_c$ will be less than or equal to our estimated $\lambda_c$, hence establishing an conservative estimate (i.e., biased away from zero) of system stability. Similarly, estimates of $\lambda$ relying on autocorrelation ($\lambda_\mathrm{AC1}$ and $\lambda_{\mathrm{Var}}$) will also be bounded due to Jensen's inequality. As the exponential function used to estimate $\lambda_{d,\mathrm{AC1}}$ is convex ($\mathrm{AC1} = \frac{1}{n} \sum_{i=1}^n e^{-\lambda_c \Delta t_i}$), our empirical AC1 estimates will be greater than or equal to $e^{-\lambda_c \langle \Delta t \rangle}$; inflation in AC1 implies a slower restoring rate and hence the true $\lambda_c$ will be greater than or equal to $\lambda_{c,\mathrm{AC1}}$ and $\lambda_{c,\mathrm{Var}}$ (i.e., less conservative, as $\lambda$ is biased towards zero). 

In both cases, the accuracy of the $\lambda_c$ estimate will be directly related to the variance of the sampling intervals; the closer the individual $\Delta t_i$ values are to being strictly equal, the closer the estimators approach the true $\lambda_c$. We hence use $\langle \Delta t \rangle$ as an analytical correction for each $\lambda_d$ estimate to bring them into a common physical reference frame and as close to the true continuous-time $\lambda_c$ as is possible with discretely-sampled data, despite the fact that $\lambda_{\mathrm{AC1}}$ and $\lambda_{\mathrm{Var}}$ are not strictly valid for unevenly sampled data and $\langle \Delta t \rangle$ is not a perfect solution. We note that a regression-based $\lambda$ estimate maintains the core advantage over $\lambda_{\mathrm{AC1}}$ and $\lambda_{\mathrm{Var}}$ that it integrates $\Delta t_i$ directly into each point-by-point estimate prior to regression for $\lambda$, which minimizes the influence of irregular time sampling and removes the core requirement that data be evenly sampled in time to estimate $\lambda$. As resampling and temporal interpolation techniques can bias, e.g., AC1 and variance, a regression-based $\lambda$ formulation is preferred for unevenly sampled data. 

\subsection*{Deseasoning and Detrending}

For our synthetic (Figure \ref{f1}, Supplemental Figure S2) and vegetation (Supplemental Figure S3) analyses, we compare our regression-based $\lambda$ to that recovered from standard data pre-processing. In short, we first deseason and detrend each time series using (1) Seasonal Trend Decomposition via Loess (STL) \citep{STL}, (2) by removing both the long-term monthly mean and a five-year rolling average (here termed climatological deseasoning), and (3) using a harmonic fit for seasonality followed by a five-year rolling average \citep{smith2023b}. All three approaches yield nominally deseasoned and detrended residual time series that can be used to compute $\lambda$ via autocorrelation ($\lambda_{\mathrm{AC1}}$) or variance ($\lambda_{\mathrm{Var}}$) \citep{smith2022}. For STL, we use a seasonal period of 365 days and a smoothing window of seven, as has been used in several previous publications \citep{smith2023b}.  

\subsection*{Synthetic Time Series Data}

We generate a one-dimensional time series describing a pitchfork-bifurcation model with seasonal forcing \citep{smith2025} to test our regression-based $\lambda$ estimate and compare it to common data analysis workflows: 

\begin{equation}
\frac{\dint x}{\dint t} = p(t)\,x - x^3 - \lambda x + A \cos(\omega t) + \sigma \,\xi(t)
\end{equation}

\noindent We vary the control parameter $p$ linearly to produce the transition from a high to a low state (Figure \ref{f1}). Seasonality is controlled by the amplitude $A$ and periodicity $\cos(\omega t)$, with additive Gaussian white noise $\sigma \xi(t)$. We use a model run with seasonal amplitude $A=0$ to assess our design matrix approach to handling seasonality, as well as to provide a baseline against which to compare deseasoned and detrended seasonal data. A complete description of the model parameters used here can be found in the Supplement.

We further generate gappy versions of our time series by removing a given percentage of points randomly sampled throughout the time series (Figure \ref{f1}) or concentrated in certain periods to mimic, e.g., seasonal cloud or snow cover gaps (Supplemental Figure S1). We find that in both cases, our regression-based $\lambda$ estimates are robust to gap percentages and lengths, as compared to a gap-free control run.

\subsection*{Spatial Vegetation Model}

To explore whether our method can be adapted to the spatial case, we use the same model presented in Smith et al. \citep{smith2025}, which adapts a reaction-diffusion model \citep{dakos2010} to include seasonality. The model is parameterized for vegetation $V$, growth rate $r_v$, environmental stress $E$, diffusion coefficient $D$ and noise amplitude $\sigma$ as: 

\begin{equation}
\frac{\partial V}{\partial t} = r_v(x, y, t)\, V \left[ 1 - V \frac{(h_E^p + E^p)}{h_E^p} \right] + D \nabla^2 V + \sigma \, \eta(x, y, t)
\end{equation}

\noindent We use spatially and temporally uncorrelated Gaussian noise for $\eta(x,y,t)$. The rate and shape of the induced critical transition are controlled by $h_E$, $h_v$, and $p$ which modify the environmental stress $E$:

\begin{equation}
E = E_0(t) \, \frac{h_v}{h_v + V}
\end{equation}

\noindent We add a seasonal component to $E$ which is homogeneous across all grid cells with amplitude $A_{E_0}$:
	
\begin{equation}
E_0(t) = E_{0, \mathrm{base}}(t) + A_{E_0} \sin \left( \frac{2 \pi t}{T_{\mathrm{year}}} \right)
\end{equation}

\noindent The growth rate $r_v$ is given a seasonal component which varies in space to follow, e.g., microclimates or topographic and rainfall gradients. We thus add variability to $r_v$ pixelwise as: 
	
\begin{equation}
r_v(x, y, t) = r_{v,0} + A_{r_v}(x, y) \, \sin \!\left( \frac{2 \pi t}{T_{\mathrm{year}}} + \phi(x, y) \right)
\end{equation}

\subsubsection*{Working with Spatio-Temporal Data}
As our method of estimating $\lambda$ is regression-based, it can be simply reconfigured to analyze spatial data by collapsing the spatial dimension; in short, we can perform our regression to discover $\lambda$ using the set of all data points in a given temporal window, regardless of their spatial position. The use of spatial data can dramatically increase the data density available (e.g., 50 x 50 points per time slice instead of a single point in the one-dimensional case). This increases the statistical power of the estimated $\lambda$, and can motivate shorter temporal windows due to increased data density. For seasonal data, however, a minimal one-year window is required in order to properly capture the seasonal oscillation in the design matrix, with more data-years being preferred to better separate the seasonal attractor $\mu$ from changes in $\lambda$.

If we consider a simple spatially-explicit vegetation model, we can calculate $\frac{\Delta x_i}{\Delta t_i}$ pixel-wise so that we do not need to assume one constant seasonality or mean state for all spatial points but can rather preserve pixel-specific equilibria. We test our method on a non-seasonal control run, a model with seasonality, and two different means of deseasoning/detrending which are typically used \citep{smith2022} in the estimation of resilience on spatio-temporal data (Supplemental Figure S2). For our harmonic design matrix, we assume that spatial heterogeneity is fixed and can vary across space -- e.g., a mountain range with pockets of high- and low-biomass areas -- so that every pixel can have its own baseline equilibrium state; $\lambda$ is then measured relative to that set of local equilibria. This assumption could be violated for some systems (e.g., pattern-forming vegetation), which would require a different approach to modeling the moving seasonal attractor $\mu$ through time. 

\subsection*{Vegetation Data}

To analyze global vegetation resilience, we rely on the widely-used kernel Normalized Difference Vegetation Index (kNDVI) \citep{CampsValls2021} based on 16-day MODIS data (MOD13A2, 2000-2025 \citep{Didan2021}). We only retain the highest-quality flagged data, and resample the native 1 km data to 5 km by the spatial mean for processing speed and to compare to previous publications \citep{smith2023b}. We further remove outliers (kNDVI \textless 0.05 or \textgreater 1), and retain only points with at least 25\% data coverage; we do not gap-fill or interpolate our data. We mask our vegetation grids using MODIS land cover data (MCD12Q1, 2001-2024 \citep{modlc}), first removing non-vegetated areas. In a second step, we mask out areas in which land cover has changed during the study period (e.g., forest to grassland), in order to focus our analysis on relatively stable vegetated ecosystems \citep{smith2023b}. We split our data into two periods (2001-2012 and 2013-2025) and compute a single $\lambda$ estimate for each (Figure \ref{f2}a). We further compute the difference between these time periods (Figure \ref{f2}b) and spatially filter them (Figure \ref{f2}c) using a 5 x 5 pixel window. Within each window, we mask the center pixel unless 75\% of the pixels have the same difference sign and at least 50\% are valid. This is done to remove regions where $\lambda$ differences vary significantly in space, indicating a region of poorly-constrained $\lambda$ estimates.

In order to compare typical (i.e., AC1- and variance-based) approaches to our regression-based $\lambda$, we first detrend and deseason the vegetation data and then compute $\lambda_{\mathrm{AC1}}$ and $\lambda_{\mathrm{Var}}$ on those stationary time series. A comparison of the mean $\lambda$ estimates using different approaches can be seen in Supplemental Figure S3.

\subsection*{Paleoclimate Data}

We use NGRIP 2.5 and 5 cm sampled data \citep{NGRIP,Gkinis2014} obtained from NGRIP-1 and NGRIP-2, and use the GICC05 (yr b2k) ages included with each data set. We do not resample or otherwise pre-process the data. We compute $\lambda$ using a 200-year moving window, as was suggested by \citep{Boers2018}. We compare our $\lambda$ estimates to the timing of DO events from \citep{Boers2018}, using the main onset age (i.e., oldest date) to match the start of a DO event defined by abrupt warming at the bottom of the interstadial interval. We incorporate a simple linear drift term in our design matrix:

\begin{equation}
\frac{\Delta x_i}{\Delta t_i} = \underbrace{-\lambda}_{\beta_1} x_i + \underbrace{\lambda b}_{\beta_2} t_i + \underbrace{(\lambda a + b)}_{\beta_0} + \epsilon_i
\label{linear}
\end{equation}

\noindent as the NGRIP data does not resolve seasonality and is too short to incorporate, e.g., Milankovich cycles. As precise $\delta^{18}\text{O}$ estimates can sometimes yield identical neighboring measurements, we incorporate a small instrument uncertainty ($\pm 0.1$\textperthousand) \citep{NGRIP} to maintain the stability of weighted least squares against instances of $\frac{\Delta x_i}{\Delta t_i}=0$, especially over sections with short depth intervals where $\Delta x_i$ can be dominated by instrument noise. 

The ice core age model (GICC05) incorporates increasing uncertainties as deeper sections of the core are studied \citep{NGRIP,Boers2017}. To account for this, we use the Maximum Counting Error (MCE) provided with the NGRIP data to estimate fractional timing errors $f \approx \frac{\text{MCE}}{|t|}$ for each sliding window. As $\lambda$ is expressed as a rate, fractional uncertainty in the time domain $t$ (i.e., a stretch in $\Delta t_i$) is propagated directly into uncertainty in $\lambda$. We hence use the window-averaged mean fractional error $\bar{f} = \frac{\overline{\text{MCE}}}{|\bar{t}|}$ and expand the uncertainty envelope of $\lambda$ as $\sigma_{\lambda, \text{time}} = |\lambda| \bar{f}$. We estimate the total standard error $SE_{\lambda}$ by combining the statistical fitting error $SE_{\text{fit}}$ (from the weighted least squares regression) and the age-model error as:

\begin{equation}
SE_{\lambda} = \sqrt{ SE_{\text{fit}}^2 + \left( |\lambda| \frac{\overline{\text{MCE}}}{|\bar{t}|} \right)^2 }
\end{equation}

\noindent With this approach, we expand the $\lambda$ uncertainty envelopes of deep sections of the core where chronologies are less well-constrained, taking into account the layer counting (MCE) uncertainties relative to the GICC05 ages. 

\subsection*{Glacier Velocity Data}

We use ITS\_LIVE v2 glacier velocity data \citep{gardner2018,gardner2025} sampled over a test glacier with known surging \citep{kaab2023} in the Karakoram \citep{smith2025} (RGI2000-v7.0-G-13-05693, [71.907, 38.837]) \citep{rgi} (Figure \ref{f4}). We exclude data from before the Landsat 8 era in order to yield a relatively dense time series of velocity estimates. We test both a single point in a known surging region (Figure \ref{f4}a) and the collection of all glacier centerline points, sampled every 500 m (Figure \ref{f4}b). We do not resample the given surface velocity data to even time steps, as has been done in previous work \citep{smith2025,ticoi}, as our method natively accounts for irregular time sampling. We rely on harmonic terms in our design matrix to account for seasonal oscillations in glacier velocity -- as well as an additional quadratic term to account for non-linear acceleration -- and compute $\lambda$ using a three-year moving window (Figure \ref{f4}). 

ITS\_LIVE glacier velocity data are defined over overlapping time spans, and provide velocity ($v_i$) estimates which are averages over those observation intervals of length $S_i$. Each velocity estimate $v_i$ can hence be written as an average instantaneous velocity $u(t)$ over an observation span: 

\begin{equation}
v_i = \frac{1}{S_i} \int_{t_i - S_i/2}^{t_i + S_i/2} u(t)\,dt
\end{equation}

\noindent where $t_i$ is the midpoint of the span and $S_i$ its duration. Directly computing $\frac{\Delta v_i}{\Delta t_i}$ using fixed points (e.g., midpoints of the spans) can thus inflate derivatives when the spans overlap. We therefore incorporate the span interval $S_i$ in our $\lambda$ estimates directly; we maintain the span midpoints for gap filtering and constructing the harmonic design matrix. To integrate the variable spans $S_i$ into our $\lambda$ estimates, we define $\Delta t_{f,i}$ based on the total temporal footprint of consecutive pairs: 

\begin{equation}
\Delta t_{f,i} = \max(t_{i+1} + S_{i+1}/2,\; t_i + S_i/2) - \min(t_{i+1} - S_{i+1}/2,\; t_i - S_i/2)
\end{equation}

\noindent The span-aware derivative is thus: 

\begin{equation}
\frac{\Delta v_i}{\Delta t_{f,i}} = \frac{v_{i+1} - v_i}{\Delta t_{f,i}}
\end{equation}

\noindent and instrument error is propagated over that footprint as: 

\begin{equation}
\sigma_{\frac{\Delta v_i}{\Delta t_{f,i}}} = \frac{\sqrt{\sigma_{v_i}^2 + \sigma_{v_{i+1}}^2}}{\Delta t_{f,i}}
\end{equation}

\noindent where $\sigma_{v_i}$ is the uncertainty provided for each span-averaged glacier velocity estimate. 

The overlapping span footprint $\Delta t_{f,i}$ will always be larger than the midpoint gap between sequential observations ($\Delta t_{m,i}$), which serves to suppress or even eliminate derivative inflation for closely-spaced span midpoints. For example, two neighboring spans of 10 and 100 days with a one-day midpoint separation would yield a small $\Delta t_{m,i}$ (1 day) and a much larger span-aware $\Delta t_{f,i}$ (105 days); for the same $\Delta v_i$, the derivatives would vary greatly. Hence, the span-aware $\lambda$ derived from $\frac{\Delta v_i}{\Delta t_{f,i}}$ is attenuated when compared to the naive estimate ($\frac{\Delta v_i}{\Delta t_{m,i}}$), with the factor difference being approximately $\frac{\Delta t_{f,i}}{\Delta t_{m,i}}$. Both metrics, however, can be used as statistical early-warning signals of oncoming surges; they differ in how strictly they adhere to the underlying mathematical framework of Langevin dynamics. The span-aware $\lambda$ is more physically consistent, while the naive midpoint $\lambda$ is conceptually simpler at the cost of inflating $\lambda$ estimates in proportion to the span overlap in the underlying data. Further, they yield different outlier distributions -- the span-aware $\lambda$ tends to have many fewer outliers as it corrects for the derivative inflation of the midpoint-based approach. This does not make a substantial difference for the 1D case, as outliers are handled cleanly by our robust linear solver. For the 2D case, however, large coherent blocks of outliers (due to, e.g., many inflated derivatives over the entire glacier coherently in time) can mislead our robust solver and yield noise-dominated $\lambda$ estimates. To minimize potential biases from these inflated derivatives, we rely on a span-aware $\lambda$ estimate in both the 1D and 2D cases (Figure \ref{f4}).

\section*{Data Availability}

All data is publicly available. Vegetation data was accessed via Google Earth Engine \citep{gorelick2017}; the original data can be found here: \url{https://doi.org/10.5067/MODIS/MOD13A2.061} \citep{Didan2021}. NGRIP ice core data can be found here: \url{https://www.iceandclimate.nbi.ku.dk/data/} \citep{NGRIP,Gkinis2014}. ITS\_LIVE glacier velocity data was accessed via the public Python API \citep{gardner2018,gardner2025}. 

\section*{Code Availability}

Synthetic data creation, glacier velocity data access, and analysis scripts are publicly available on Zenodo: \url{https://doi.org/10.5281/zenodo.19731234}.

\section*{Acknowledgments}
T.S. acknowledges support from the DFG STRIVE project (SM 710/2-1) and the Universit\"{a}t Potsdam Remote Sensing Computational Cluster. This is ClimTip contribution \#152; the ClimTip project has received funding from the European Union's Horizon Europe research and innovation programme under grant agreement No. 101137601. N.B. acknowledges additional funding by the Volkswagen Foundation, the European Space Agency Climate Change Initiative (ESA-CCI) Tipping Elements SIRENE project (contract no. 4000146954/24/I-LR), and the Past to Future (P2F) project, which has received funding from the European Union’s Horizon Europe research and innovation programme under grant agreement No. 101184070.

\section*{Author Contributions}
T.S. conceived and designed the study, processed the data, and performed the numerical analysis. T.S. wrote the manuscript with contributions from A.M., C.S., and N.B.

\section*{Competing Interests}
The authors declare no competing interests.

\clearpage
\newpage
\singlespacing

\clearpage
\newpage

\pagenumbering{gobble}

\begin{centering}
\clearpage\thispagestyle{empty}
\textbf{\Large{Supplement to: Estimating the Resilience of Non-Stationary Systems}} 
\vspace{2cm}

Taylor Smith$^{1}$*, Andreas Morr$^{2,3}$, Christof Sch\"{o}tz$^{3,4}$, Niklas Boers$^{3,4}$\\
$^{1}$Institute of Geosciences, Universit\"{a}t Potsdam, Potsdam, Germany\\
$^{2}$Department of Mathematics, School of Computation, Information and Technology, Technical University of Munich, Munich, Germany  \\
$^{3}$Potsdam Institute for Climate Impact Research, Potsdam, Germany\\
$^{4}$Munich Climate Center and Earth System Modelling Group, Department of Aerospace and Geodesy, TUM School of Engineering and Design, Technical University of Munich, Munich, Germany  \\

\end{centering}

\vspace{15cm}

\noindent
Corresponding author: \\
Taylor Smith \\
Email: tasmith@uni-potsdam.de

\clearpage
\doublespacing

\setcounter{figure}{0}
\renewcommand{\figurename}{Supplementary Figure}
\renewcommand{\thefigure}{S\arabic{figure}}
\renewcommand{\thetable}{S\arabic{table}}

\clearpage
\newpage

\section*{Synthetic Time Series Model}

For the time series model, we use the same model setup as was used in Smith et al. (2026). Scripts to reproduce our model code can be found on Zenodo: \\
\noindent\textit{T Smith. (2025). Predicting Instabilities in Transient Landforms and Interconnected Ecosystems. Zenodo. https://doi.org/10.5281/zenodo.18031340}

The parameters used can also be found here:

\begin{table}[h!]
	\centering
	\begin{tabular}{ll}
		\toprule
		\textbf{Parameter} & \textbf{Value / Description} \\
		\midrule
		Time step $\Delta t$ & $0.01$~days \\
		Simulation duration & 1950–2025 \\
		Initial condition $x_0$ & $0.01$ \\
		Control parameter range ($p_{\mathrm{start}}, p_{\mathrm{end}}$) & $(-0.5, 0.5)$ \\
		Linear damping coefficient & $0.0$ \\
		Noise amplitude $\sigma$ & $0.025$ \\
		Seasonal amplitude $A_{\mathrm{seasonal}}$ & $0.05$ \\
		Seasonal frequency $\omega$ & 365~day \\
		Integration scheme & Euler–Maruyama \\
		Rescaling range (output) & $[-0.1, 1]$ \\
		\bottomrule
	\end{tabular}
	\caption{Parameters used in the pitchfork bifurcation time series model.}
\end{table}

\section*{Gap Robustness}

\begin{figure*}[!h]
\centering
\includegraphics[width=0.75\linewidth]{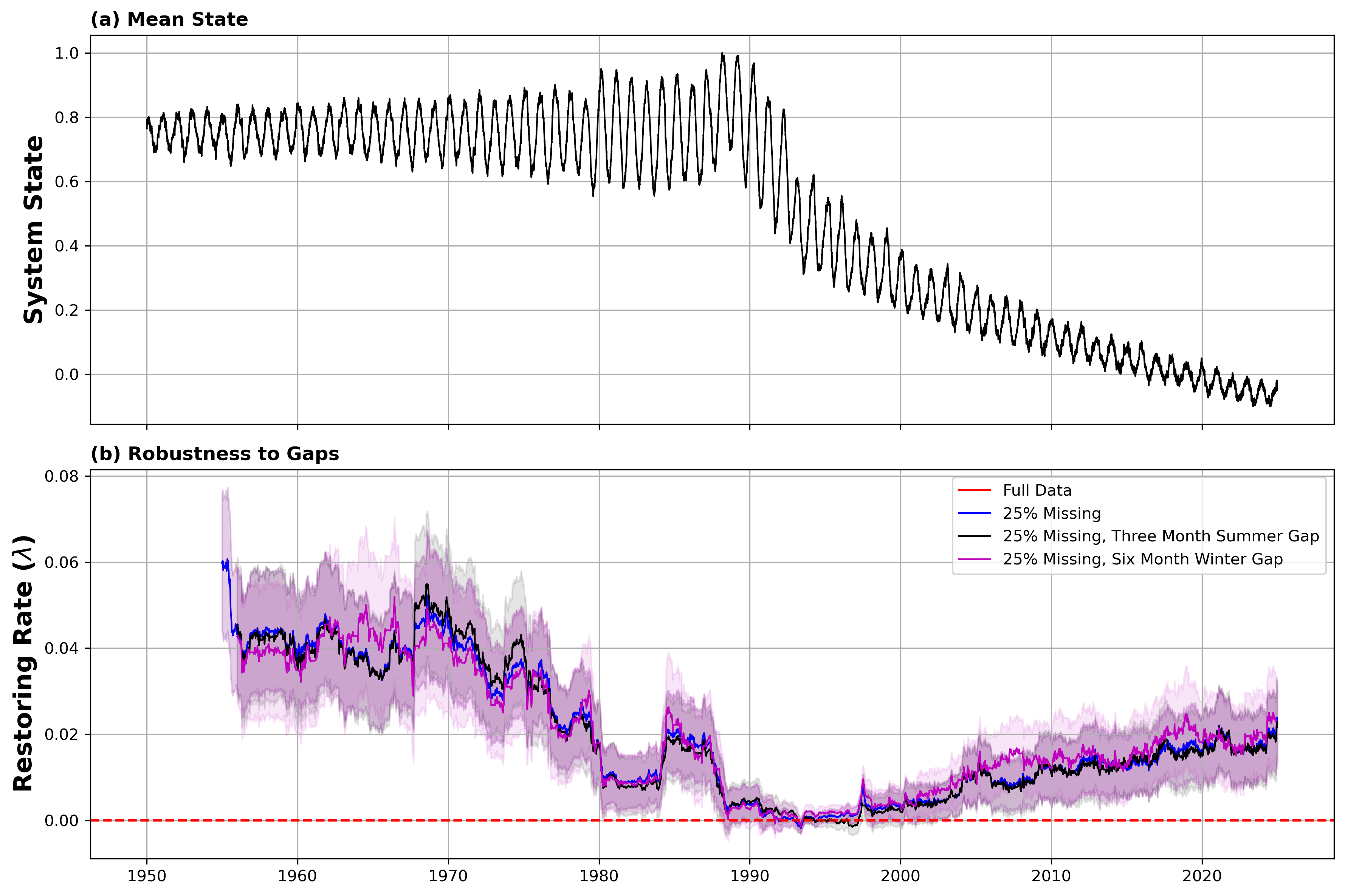}
\caption{\textbf{Gab Robustness.} (A) System state. (B) Restoring rate $\lambda$ estimated on data sets with variable gap percentages and gap lengths.}
\end{figure*}

\section*{Synthetic Vegetation Model}

For the reaction-diffusion spatial vegetation model, we use the same model setup as was used in Smith et al. (2026). Scripts to reproduce our model code can be found on Zenodo: \\
\noindent\textit{T Smith. (2025). Predicting Instabilities in Transient Landforms and Interconnected Ecosystems. Zenodo. https://doi.org/10.5281/zenodo.18031340}

The parameters used can also be found here:

\begin{table}[h!]
	\centering
	\begin{tabular}{ll}
		\toprule
		\textbf{Parameter} & \textbf{Value / Description} \\
		\midrule
		Grid size & $50 \times 50$ \\
		Time step $\Delta t$ & $0.1$~days \\
		Simulation duration & $20$~years \\
		Base growth rate $r_v$ & $0.5$ \\
		Vegetation half-saturation constant $h_v$ & $0.2$ \\
		Environmental half-saturation constant $h_E$ & $2.0$ \\
		Nonlinearity exponent $p$ & $4.0$ \\
		Diffusion coefficient $D$ & $0.5$ \\
		Noise amplitude $\sigma$ & $0.02$ \\
		Initial environmental forcing $E_{0,\mathrm{start}}$ & $4.0$ \\
		Final environmental forcing $E_{0,\mathrm{end}}$ & $8.0$ \\
		Seasonal amplitude in $E_0$ ($A_{E_0}$) & $0.15$ \\
		Mean seasonal amplitude in $r_v$ ($A_{r_v,\mathrm{mean}}$) & $0.15$ \\
		Std. of amplitude heterogeneity ($A_{r_v,\mathrm{std}}$) & $0.03$ \\
		Max spatial phase shift & $\pi/6$ ($\sim$ 30 days) \\
		Seasonal period $T_{\mathrm{season}}$ & $365$~days \\
		Integration scheme & Euler–Maruyama \\
		\bottomrule
	\end{tabular}
	\caption{Parameters used in the spatial reaction-diffusion vegetation model.}
\end{table}

\begin{figure*}[!h]
\centering
\includegraphics[width=0.75\linewidth]{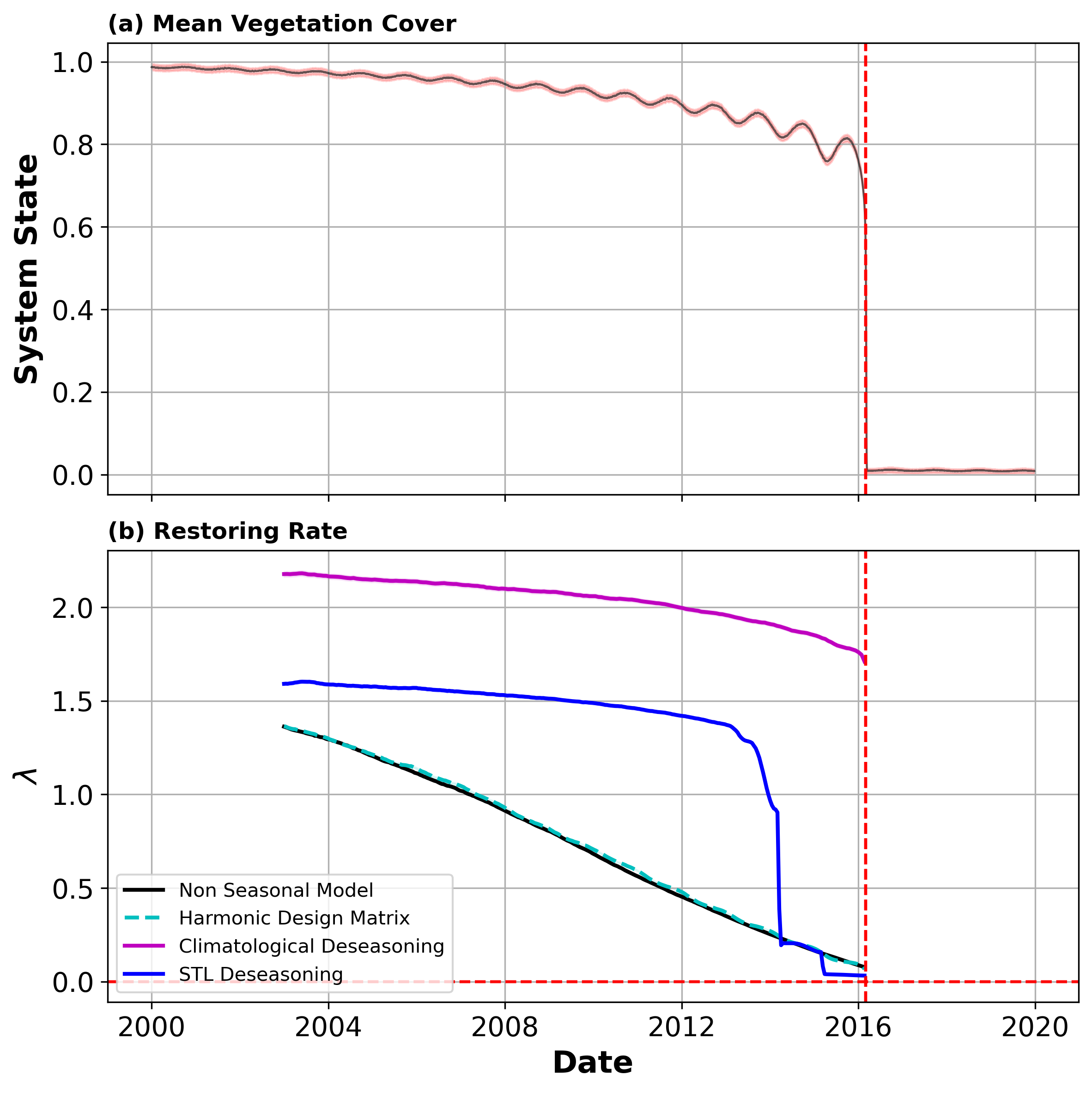}
\caption{\textbf{Stability of a spatially-extended vegetation model.} (A) Mean vegetation state. (B) Restoring rate $\lambda$ estimated on four different data processing schemes. Note that the non-seasonal control model and the seasonal model using a harmonic design matrix (Methods) yield almost identical results.} 
\end{figure*}

\clearpage
\newpage

\section*{Vegetation Data}

\begin{figure*}[!h]
\centering
\includegraphics[width=\linewidth]{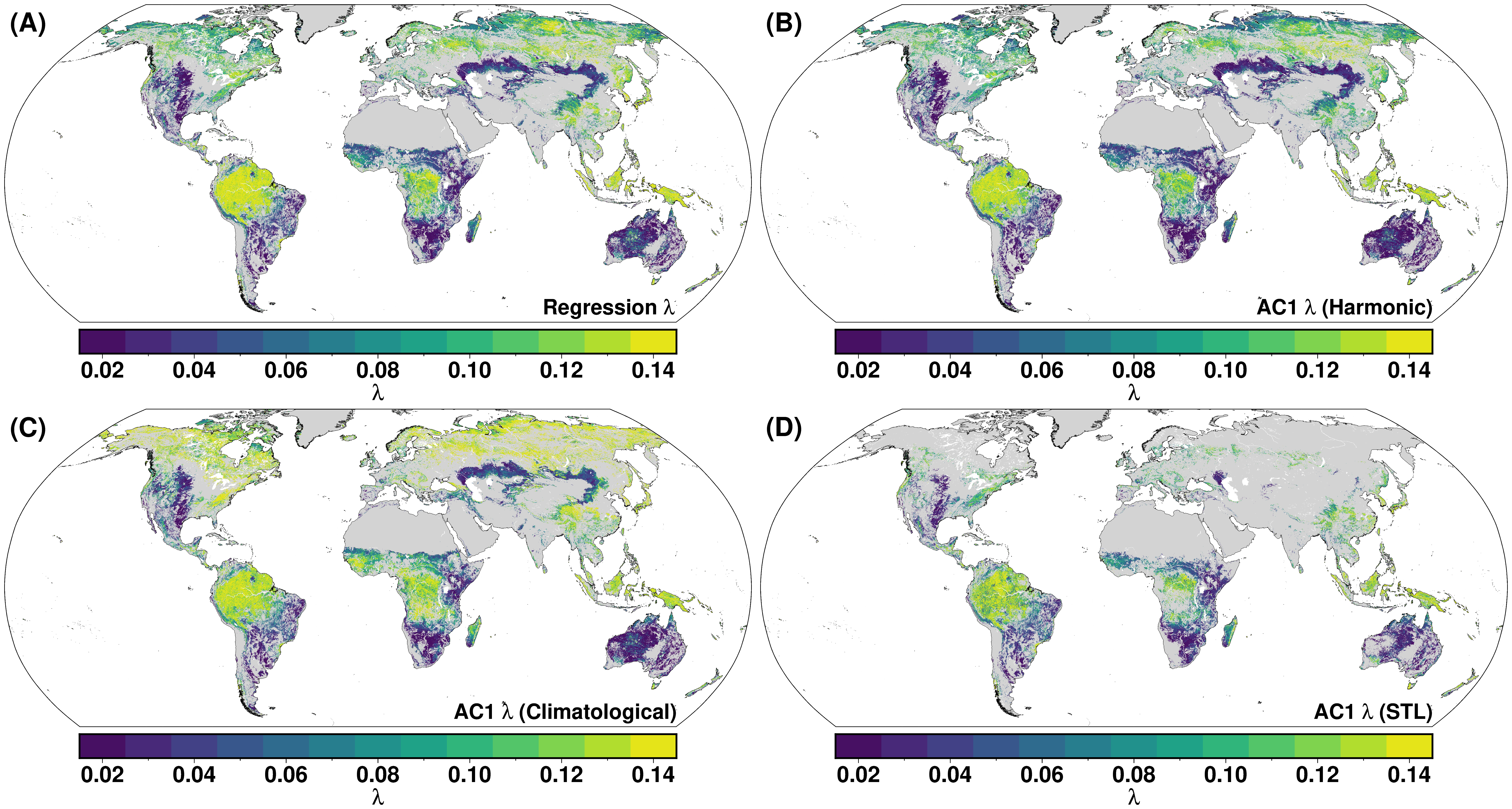}
\caption{\textbf{Global Vegetation Data.} (A) Regression-based $\lambda$ using a harmonic-linear design matrix. (B,C,D) $\lambda_\mathrm{AC1}$ estimated on deseasoned and detrended data: (B) harmonic deseasoning and rolling mean detrending, (C) climatological deseasoning and rolling mean detrending, (D) STL deseasoning and detrending.}
\end{figure*}

\clearpage
\newpage

\section*{Paleoclimate Data}

\begin{figure*}[!h]
\centering
\includegraphics[width=\linewidth]{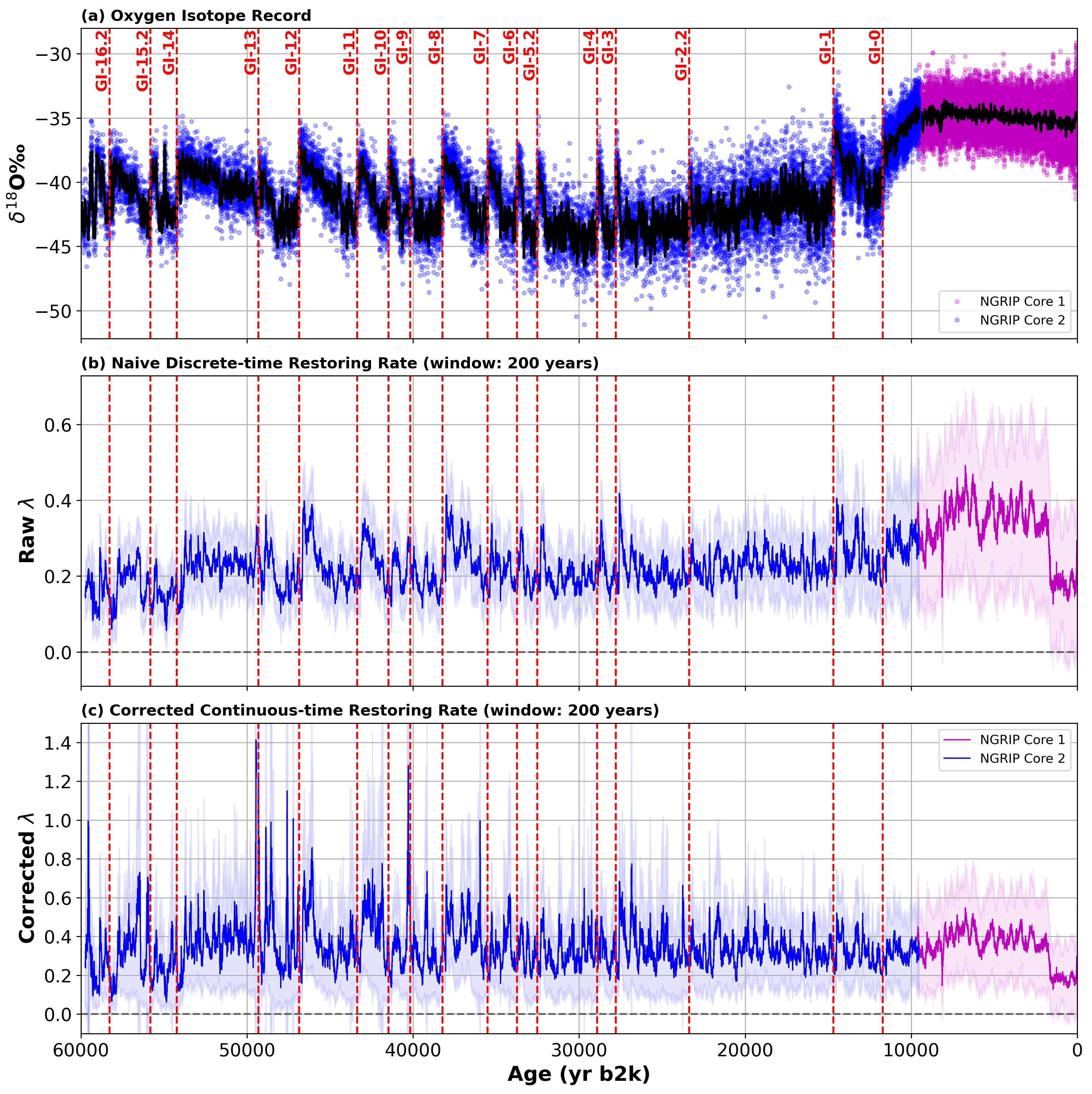}
\caption{\textbf{NGRIP Ice Core Data.} (A) $\delta^{18}\text{O}$ isotope records from NGRIP. Vertical lines mark DO event timing after Boers (2018). (B) Changes in $\lambda$ calculated on the raw time series without pre-processing, incorporating measurement and age-model uncertainties via weighted least squares (Methods). (C) Corrected $\lambda$ accounting for changes in sampling intervals through time (Methods), showing less pronounced decreases in $\lambda$ before many DO events.}
\end{figure*}

\clearpage
\newpage

\section*{Methods}

\begin{figure*}[!h]
\centering
\includegraphics[width=\linewidth]{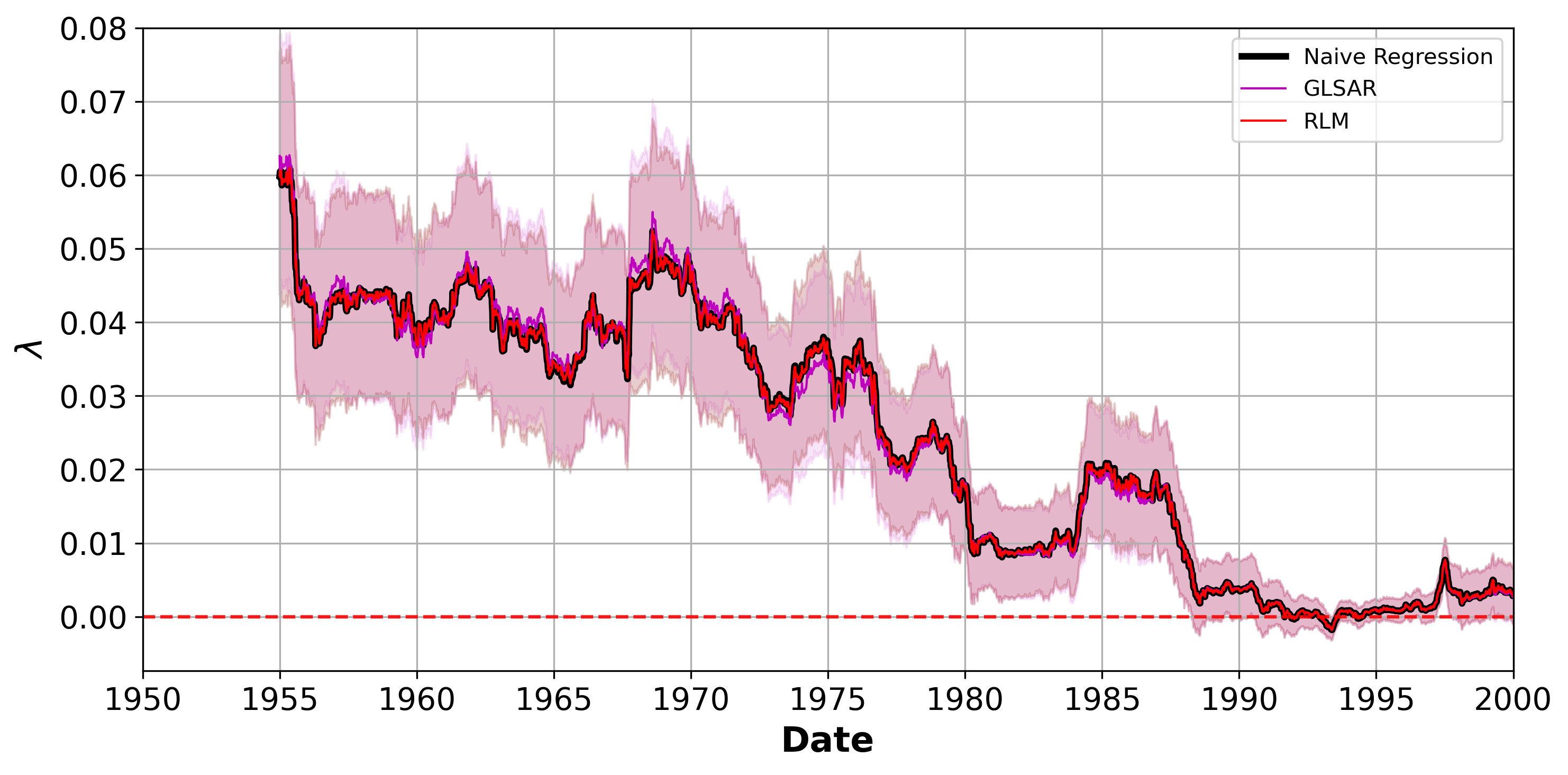}
\caption{\textbf{Comparison of different regression models used to recover $\lambda$}. Naive ordinary least squares (black), Generalized least squares (purple) and robust regression (red) on the same model used in Figure 1. Depending on the system under study, some regression approaches may be more or less appropriate; our method allows for a flexible choice of regression framework.}
\end{figure*}


\begin{thebibliography}{10}
	\expandafter\ifx\csname url\endcsname\relax
	\def\url#1{\texttt{#1}}\fi
	\expandafter\ifx\csname urlprefix\endcsname\relax\def\urlprefix{URL }\fi
	\providecommand{\bibinfo}[2]{#2}
	\providecommand{\eprint}[2][]{\url{#2}}
	
	\bibitem{Lenton2008}
	\bibinfo{author}{Lenton, T.~M.} \emph{et~al.}
	\newblock \bibinfo{title}{{Tipping elements in the Earth's climate system}}.
	\newblock \emph{\bibinfo{journal}{Proceedings of the National Academy of
			Sciences}} \textbf{\bibinfo{volume}{105}}, \bibinfo{pages}{1786--1793}
	(\bibinfo{year}{2008}).
	\newblock \urlprefix\url{http://www.pnas.org/content/105/6/1786}.
	
	\bibitem{boers2022_erl}
	\bibinfo{author}{Boers, N.}, \bibinfo{author}{Ghil, M.} \&
	\bibinfo{author}{Stocker, T.~F.}
	\newblock \bibinfo{title}{Theoretical and paleoclimatic evidence for abrupt
		transitions in the earth system}.
	\newblock \emph{\bibinfo{journal}{Environmental Research Letters}}
	\textbf{\bibinfo{volume}{17}}, \bibinfo{pages}{093006}
	(\bibinfo{year}{2022}).
	\newblock \urlprefix\url{https://dx.doi.org/10.1088/1748-9326/ac8944}.
	
	\bibitem{lenton2024}
	\bibinfo{author}{Lenton, T.~M.} \emph{et~al.}
	\newblock \bibinfo{title}{Remotely sensing potential climate change tipping
		points across scales}.
	\newblock \emph{\bibinfo{journal}{Nature Communications}}
	\textbf{\bibinfo{volume}{15}}, \bibinfo{pages}{343} (\bibinfo{year}{2024}).
	\newblock \urlprefix\url{https://doi.org/10.1038/s41467-023-44609-w}.
	
	\bibitem{forzieri2022}
	\bibinfo{author}{Forzieri, G.}, \bibinfo{author}{Dakos, V.},
	\bibinfo{author}{McDowell, N.~G.}, \bibinfo{author}{Ramdane, A.} \&
	\bibinfo{author}{Cescatti, A.}
	\newblock \bibinfo{title}{Emerging signals of declining forest resilience under
		climate change}.
	\newblock \emph{\bibinfo{journal}{Nature}} \bibinfo{pages}{1--6}
	(\bibinfo{year}{2022}).
	\newblock \urlprefix\url{https://doi.org/10.1038/s41586-022-04959-9}.
	
	\bibitem{boers2021}
	\bibinfo{author}{Boers, N.} \& \bibinfo{author}{Rypdal, M.}
	\newblock \bibinfo{title}{Critical slowing down suggests that the western
		greenland ice sheet is close to a tipping point}.
	\newblock \emph{\bibinfo{journal}{Proceedings of the National Academy of
			Sciences}} \textbf{\bibinfo{volume}{118}}, \bibinfo{pages}{e2024192118}
	(\bibinfo{year}{2021}).
	\newblock \urlprefix\url{https://doi.org/10.1073/pnas.2024192118}.
	
	\bibitem{boers2021b}
	\bibinfo{author}{Boers, N.}
	\newblock \bibinfo{title}{Observation-based early-warning signals for a
		collapse of the atlantic meridional overturning circulation}.
	\newblock \emph{\bibinfo{journal}{Nature Climate Change}}
	\textbf{\bibinfo{volume}{11}}, \bibinfo{pages}{680--688}
	(\bibinfo{year}{2021}).
	\newblock \urlprefix\url{https://doi.org/10.1038/s41558-021-01097-4}.
	
	\bibitem{boers2025}
	\bibinfo{author}{Boers, N.} \emph{et~al.}
	\newblock \bibinfo{title}{Destabilization of earth system tipping elements}.
	\newblock \emph{\bibinfo{journal}{Nature Geoscience}}
	\textbf{\bibinfo{volume}{18}}, \bibinfo{pages}{949--960}
	(\bibinfo{year}{2025}).
	
	\bibitem{Carpenter2006}
	\bibinfo{author}{Carpenter, S.~R.} \& \bibinfo{author}{Brock, W.~a.}
	\newblock \bibinfo{title}{{Rising variance: a leading indicator of ecological
			transition.}}
	\newblock \emph{\bibinfo{journal}{Ecology letters}}
	\textbf{\bibinfo{volume}{9}}, \bibinfo{pages}{311--8} (\bibinfo{year}{2006}).
	\newblock \urlprefix\url{http://www.ncbi.nlm.nih.gov/pubmed/16958897}.
	
	\bibitem{Dakos2008}
	\bibinfo{author}{Dakos, V.} \emph{et~al.}
	\newblock \bibinfo{title}{{Slowing down as an early warning signal for abrupt
			climate change}}.
	\newblock \emph{\bibinfo{journal}{Proceedings of the National Academy of
			Sciences of the United States of America}} \textbf{\bibinfo{volume}{105}},
	\bibinfo{pages}{14308--12} (\bibinfo{year}{2008}).
	\newblock \urlprefix\url{https://doi.org/10.1073/pnas.0802430105}.
	
	\bibitem{Scheffer2009}
	\bibinfo{author}{Scheffer, M.} \emph{et~al.}
	\newblock \bibinfo{title}{{Early-warning signals for critical transitions}}.
	\newblock \emph{\bibinfo{journal}{Nature}} \textbf{\bibinfo{volume}{461}},
	\bibinfo{pages}{53--9} (\bibinfo{year}{2009}).
	\newblock \urlprefix\url{http://www.ncbi.nlm.nih.gov/pubmed/19727193}.
	
	\bibitem{dakos2010}
	\bibinfo{author}{Dakos, V.}, \bibinfo{author}{van Nes, E.~H.},
	\bibinfo{author}{Donangelo, R.}, \bibinfo{author}{Fort, H.} \&
	\bibinfo{author}{Scheffer, M.}
	\newblock \bibinfo{title}{Spatial correlation as leading indicator of
		catastrophic shifts}.
	\newblock \emph{\bibinfo{journal}{Theoretical Ecology}}
	\textbf{\bibinfo{volume}{3}}, \bibinfo{pages}{163--174}
	(\bibinfo{year}{2010}).
	\newblock \urlprefix\url{https://doi.org/10.1007/s12080-009-0060-6}.
	
	\bibitem{smith2022}
	\bibinfo{author}{Smith, T.}, \bibinfo{author}{Traxl, D.} \&
	\bibinfo{author}{Boers, N.}
	\newblock \bibinfo{title}{Empirical evidence for recent global shifts in
		vegetation resilience}.
	\newblock \emph{\bibinfo{journal}{Nature Climate Change}}
	\textbf{\bibinfo{volume}{12}}, \bibinfo{pages}{477--484}
	(\bibinfo{year}{2022}).
	\newblock \urlprefix\url{https://doi.org/10.1038/s41558-022-01352-2}.
	
	\bibitem{smith2025}
	\bibinfo{author}{Smith, T.}, \bibinfo{author}{Morr, A.},
	\bibinfo{author}{Bookhagen, B.} \& \bibinfo{author}{Boers, N.}
	\newblock \bibinfo{title}{Predicting instabilities in transient landforms and
		interconnected ecosystems} (\bibinfo{year}{2025}).
	
	\bibitem{grziwotz2023}
	\bibinfo{author}{Grziwotz, F.} \emph{et~al.}
	\newblock \bibinfo{title}{Anticipating the occurrence and type of critical
		transitions}.
	\newblock \emph{\bibinfo{journal}{Science Advances}}
	\textbf{\bibinfo{volume}{9}}, \bibinfo{pages}{eabq4558}
	(\bibinfo{year}{2023}).
	\newblock
	\urlprefix\url{https://www.science.org/doi/abs/10.1126/sciadv.abq4558}.
	\newblock \eprint{https://www.science.org/doi/pdf/10.1126/sciadv.abq4558}.
	
	\bibitem{morr2024b}
	\bibinfo{author}{Morr, A.}, \bibinfo{author}{Riechers, K.},
	\bibinfo{author}{Gorjão, L.~R.} \& \bibinfo{author}{Boers, N.}
	\newblock \bibinfo{title}{Anticipating critical transitions in multidimensional
		systems driven by time- and state-dependent noise}.
	\newblock \emph{\bibinfo{journal}{Physical Review Research}}
	\textbf{\bibinfo{volume}{6}}, \bibinfo{pages}{033251} (\bibinfo{year}{2024}).
	
	\bibitem{bury2020}
	\bibinfo{author}{Bury, T.~M.}, \bibinfo{author}{Bauch, C.~T.} \&
	\bibinfo{author}{Anand, M.}
	\newblock \bibinfo{title}{Detecting and distinguishing tipping points using
		spectral early warning signals}.
	\newblock \emph{\bibinfo{journal}{Journal of the Royal Society Interface}}
	\textbf{\bibinfo{volume}{17}}, \bibinfo{pages}{20200482}
	(\bibinfo{year}{2020}).
	\newblock \urlprefix\url{https://doi.org/10.1098/rsif.2020.0482}.
	
	\bibitem{morr2024}
	\bibinfo{author}{Morr, A.} \& \bibinfo{author}{Boers, N.}
	\newblock \bibinfo{title}{Detection of approaching critical transitions in
		natural systems driven by red noise}.
	\newblock \emph{\bibinfo{journal}{Physical Review X}}
	\textbf{\bibinfo{volume}{14}}, \bibinfo{pages}{021037}
	(\bibinfo{year}{2024}).
	\newblock \urlprefix\url{https://doi.org/10.1103/PhysRevX.14.021037}.
	
	\bibitem{bury2021}
	\bibinfo{author}{Bury, T.~M.} \emph{et~al.}
	\newblock \bibinfo{title}{Deep learning for early warning signals of tipping
		points}.
	\newblock \emph{\bibinfo{journal}{Proceedings of the National Academy of
			Sciences}} \textbf{\bibinfo{volume}{118}}, \bibinfo{pages}{e2106140118}
	(\bibinfo{year}{2021}).
	\newblock \urlprefix\url{https://doi.org/10.1073/pnas.2106140118}.
	
	\bibitem{huang2024}
	\bibinfo{author}{Huang, Y.}, \bibinfo{author}{Bathiany, S.},
	\bibinfo{author}{Ashwin, P.} \& \bibinfo{author}{Boers, N.}
	\newblock \bibinfo{title}{Deep learning for predicting rate-induced tipping}.
	\newblock \emph{\bibinfo{journal}{Nature Machine Intelligence}}
	\bibinfo{pages}{1--10} (\bibinfo{year}{2024}).
	\newblock \urlprefix\url{https://doi.org/10.1038/s42256-024-00937-0}.
	
	\bibitem{smith2023b}
	\bibinfo{author}{Smith, T.} \& \bibinfo{author}{Boers, N.}
	\newblock \bibinfo{title}{Reliability of vegetation resilience estimates
		depends on biomass density}.
	\newblock \emph{\bibinfo{journal}{Nature Ecology \& Evolution}}
	\textbf{\bibinfo{volume}{7}}, \bibinfo{pages}{1799--1808}
	(\bibinfo{year}{2023}).
	\newblock \urlprefix\url{https://doi.org/10.1038/s41559-023-02194-7}.
	
	\bibitem{STL}
	\bibinfo{author}{Cleveland, R.~B.}, \bibinfo{author}{Cleveland, W.~S.},
	\bibinfo{author}{McRae, J.~E.} \& \bibinfo{author}{Terpenning, I.}
	\newblock \bibinfo{title}{Stl: A seasonal-trend decomposition procedure based
		on loess}.
	\newblock \emph{\bibinfo{journal}{Journal of Official Statistics}}
	\textbf{\bibinfo{volume}{6}}, \bibinfo{pages}{3--73} (\bibinfo{year}{1990}).
	
	\bibitem{Djikstra2013}
	\bibinfo{author}{Djikstra, H.}
	\newblock \emph{\bibinfo{title}{{Nonlinear Climate Dynamics}}}
	(\bibinfo{publisher}{Cambridge University Press}, \bibinfo{address}{New
		York}, \bibinfo{year}{2013}).
	
	\bibitem{rietkerk2025}
	\bibinfo{author}{Rietkerk, M.}, \bibinfo{author}{Skiba, V.},
	\bibinfo{author}{Weinans, E.}, \bibinfo{author}{H{\'e}bert, R.} \&
	\bibinfo{author}{Laepple, T.}
	\newblock \bibinfo{title}{Ambiguity of early warning signals for climate
		tipping points}.
	\newblock \emph{\bibinfo{journal}{Nature Climate Change}}
	\bibinfo{pages}{1--10} (\bibinfo{year}{2025}).
	\newblock \urlprefix\url{https://doi.org/10.1038/s41558-025-02328-8}.
	
	\bibitem{Liu2026}
	\bibinfo{author}{Liu, T.} \emph{et~al.}
	\newblock \bibinfo{title}{Data gaps and outliers distort
		critical-slowing-down-based resilience indicators}.
	\newblock \emph{\bibinfo{journal}{Science Advances}}
	\textbf{\bibinfo{volume}{12}} (\bibinfo{year}{2026}).
	
	\bibitem{smith2022b}
	\bibinfo{author}{Smith, T.} \emph{et~al.}
	\newblock \bibinfo{title}{Reliability of resilience estimation based on
		multi-instrument time series}.
	\newblock \emph{\bibinfo{journal}{Earth System Dynamics}}
	\textbf{\bibinfo{volume}{14}}, \bibinfo{pages}{173--183}
	(\bibinfo{year}{2023}).
	\newblock \urlprefix\url{https://esd.copernicus.org/articles/14/173/2023/}.
	
	\bibitem{benyami2023}
	\bibinfo{author}{Ben-Yami, M.}, \bibinfo{author}{Skiba, V.},
	\bibinfo{author}{Bathiany, S.} \& \bibinfo{author}{Boers, N.}
	\newblock \bibinfo{title}{Uncertainties in critical slowing down indicators of
		observation-based fingerprints of the atlantic overturning circulation}.
	\newblock \emph{\bibinfo{journal}{Nature Communications}}
	\textbf{\bibinfo{volume}{14}} (\bibinfo{year}{2023}).
	
	\bibitem{benyami2024}
	\bibinfo{author}{Ben-Yami, M.}, \bibinfo{author}{Morr, A.},
	\bibinfo{author}{Bathiany, S.} \& \bibinfo{author}{Boers, N.}
	\newblock \bibinfo{title}{Uncertainties too large to predict tipping times of
		major earth system components from historical data}.
	\newblock \emph{\bibinfo{journal}{Science Advances}}
	\textbf{\bibinfo{volume}{10}} (\bibinfo{year}{2024}).
	
	\bibitem{Boers2017}
	\bibinfo{author}{Boers, N.}, \bibinfo{author}{Goswami, B.} \&
	\bibinfo{author}{Ghil, M.}
	\newblock \bibinfo{title}{A complete representation of uncertainties in
		layer-counted paleoclimatic archives}.
	\newblock \emph{\bibinfo{journal}{Climate of the Past}}
	\textbf{\bibinfo{volume}{13}}, \bibinfo{pages}{1169--1180}
	(\bibinfo{year}{2017}).
	
	\bibitem{boulton2022}
	\bibinfo{author}{Boulton, C.~A.}, \bibinfo{author}{Lenton, T.~M.} \&
	\bibinfo{author}{Boers, N.}
	\newblock \bibinfo{title}{Pronounced loss of amazon rainforest resilience since
		the early 2000s}.
	\newblock \emph{\bibinfo{journal}{Nature Climate Change}}
	\textbf{\bibinfo{volume}{12}}, \bibinfo{pages}{271--278}
	(\bibinfo{year}{2022}).
	\newblock \urlprefix\url{https://doi.org/10.1038/s41558-022-01287-8}.
	
	\bibitem{blaschke2024}
	\bibinfo{author}{Blaschke, L.~L.} \emph{et~al.}
	\newblock \bibinfo{title}{Spatial correlation increase in single-sensor
		satellite data reveals loss of amazon rainforest resilience}.
	\newblock \emph{\bibinfo{journal}{Earth's Future}}
	\textbf{\bibinfo{volume}{12}}, \bibinfo{pages}{e2023EF004040}
	(\bibinfo{year}{2024}).
	\newblock
	\urlprefix\url{https://agupubs.onlinelibrary.wiley.com/doi/abs/10.1029/2023EF004040}.
	\newblock \bibinfo{note}{E2023EF004040 2023EF004040}.
	
	\bibitem{verbesselt2016}
	\bibinfo{author}{Verbesselt, J.} \emph{et~al.}
	\newblock \bibinfo{title}{Remotely sensed resilience of tropical forests}.
	\newblock \emph{\bibinfo{journal}{Nature Climate Change}}
	\textbf{\bibinfo{volume}{6}}, \bibinfo{pages}{1028--1031}
	(\bibinfo{year}{2016}).
	\newblock \urlprefix\url{https://doi.org/10.1038/nclimate3108}.
	
	\bibitem{CampsValls2021}
	\bibinfo{author}{Camps-Valls, G.} \emph{et~al.}
	\newblock \bibinfo{title}{A unified vegetation index for quantifying the
		terrestrial biosphere}.
	\newblock \emph{\bibinfo{journal}{Science Advances}}
	\textbf{\bibinfo{volume}{7}} (\bibinfo{year}{2021}).
	
	\bibitem{Didan2021}
	\bibinfo{author}{Didan, K.}
	\newblock \bibinfo{title}{Modis/terra vegetation indices 16-day l3 global 1km
		sin grid v061} (\bibinfo{year}{2021}).
	
	\bibitem{VOD}
	\bibinfo{author}{Moesinger, L.} \emph{et~al.}
	\newblock \bibinfo{title}{The global long-term microwave vegetation optical
		depth climate archive (vodca)}.
	\newblock \emph{\bibinfo{journal}{Earth System Science Data}}
	\textbf{\bibinfo{volume}{12}}, \bibinfo{pages}{177--196}
	(\bibinfo{year}{2020}).
	\newblock \urlprefix\url{https://essd.copernicus.org/articles/12/177/2020/}.
	
	\bibitem{Hummel2025}
	\bibinfo{author}{Hummel, C.}, \bibinfo{author}{Boers, N.} \&
	\bibinfo{author}{Rypdal, M.}
	\newblock \bibinfo{title}{Inconclusive early warning signals for
		dansgaard-oeschger events across greenland ice cores}.
	\newblock \emph{\bibinfo{journal}{Earth System Dynamics}}
	\textbf{\bibinfo{volume}{16}}, \bibinfo{pages}{2035--2062}
	(\bibinfo{year}{2025}).
	
	\bibitem{Boettner2021}
	\bibinfo{author}{Boettner, C.}, \bibinfo{author}{Klinghammer, G.},
	\bibinfo{author}{Boers, N.}, \bibinfo{author}{Westerhold, T.} \&
	\bibinfo{author}{Marwan, N.}
	\newblock \bibinfo{title}{Early-warning signals for cenozoic climate
		transitions}.
	\newblock \emph{\bibinfo{journal}{Quaternary Science Reviews}}
	\textbf{\bibinfo{volume}{270}}, \bibinfo{pages}{107177}
	(\bibinfo{year}{2021}).
	
	\bibitem{Boers2018}
	\bibinfo{author}{Boers, N.}
	\newblock \bibinfo{title}{{Early-warning signals for Dansgaard-Oeschger events
			in a high-resolution ice core record}}.
	\newblock \emph{\bibinfo{journal}{Nature Communications}}
	\textbf{\bibinfo{volume}{9}} (\bibinfo{year}{2018}).
	\newblock \urlprefix\url{https://doi.org/10.1038/s41467-018-04881-7}.
	
	\bibitem{NGRIP}
	\bibinfo{author}{members, N. G. I. C.~P.}
	\newblock \bibinfo{title}{High-resolution record of northern hemisphere climate
		extending into the last interglacial period}.
	\newblock \emph{\bibinfo{journal}{Nature}} \textbf{\bibinfo{volume}{431}},
	\bibinfo{pages}{147--151} (\bibinfo{year}{2004}).
	
	\bibitem{Gkinis2014}
	\bibinfo{author}{Gkinis, V.}, \bibinfo{author}{Simonsen, S.},
	\bibinfo{author}{Buchardt, S.}, \bibinfo{author}{White, J.} \&
	\bibinfo{author}{Vinther, B.}
	\newblock \bibinfo{title}{Water isotope diffusion rates from the northgrip ice
		core for the last 16,000 years – glaciological and paleoclimatic
		implications}.
	\newblock \emph{\bibinfo{journal}{Earth and Planetary Science Letters}}
	\textbf{\bibinfo{volume}{405}}, \bibinfo{pages}{132--141}
	(\bibinfo{year}{2014}).
	
	\bibitem{Mitsui2024}
	\bibinfo{author}{Mitsui, T.} \& \bibinfo{author}{Boers, N.}
	\newblock \bibinfo{title}{Statistical precursor signals for
		dansgaard–oeschger cooling transitions}.
	\newblock \emph{\bibinfo{journal}{Climate of the Past}}
	\textbf{\bibinfo{volume}{20}}, \bibinfo{pages}{683--699}
	(\bibinfo{year}{2024}).
	
	\bibitem{boers2018b}
	\bibinfo{author}{Boers, N.}, \bibinfo{author}{Ghil, M.} \&
	\bibinfo{author}{Rousseau, D.-D.}
	\newblock \bibinfo{title}{Ocean circulation, ice shelf, and sea ice
		interactions explain dansgaard–oeschger cycles}.
	\newblock \emph{\bibinfo{journal}{Proceedings of the National Academy of
			Sciences}} \textbf{\bibinfo{volume}{115}} (\bibinfo{year}{2018}).
	
	\bibitem{kaab2023}
	\bibinfo{author}{K{\"a}{\"a}b, A.}, \bibinfo{author}{Bazilova, V.},
	\bibinfo{author}{Leclercq, P.~W.}, \bibinfo{author}{Mannerfelt, E.~S.} \&
	\bibinfo{author}{Strozzi, T.}
	\newblock \bibinfo{title}{Global clustering of recent glacier surges from radar
		backscatter data, 2017--2022}.
	\newblock \emph{\bibinfo{journal}{Journal of Glaciology}}
	\textbf{\bibinfo{volume}{69}}, \bibinfo{pages}{1515--1523}
	(\bibinfo{year}{2023}).
	
	\bibitem{guillet2022}
	\bibinfo{author}{Guillet, G.} \emph{et~al.}
	\newblock \bibinfo{title}{A regionally resolved inventory of high mountain asia
		surge-type glaciers, derived from a multi-factor remote sensing approach}.
	\newblock \emph{\bibinfo{journal}{The Cryosphere}}
	\textbf{\bibinfo{volume}{16}}, \bibinfo{pages}{603--623}
	(\bibinfo{year}{2022}).
	\newblock \urlprefix\url{https://tc.copernicus.org/articles/16/603/2022/}.
	
	\bibitem{ou2022}
	\bibinfo{author}{Ou, H.-W.}
	\newblock \bibinfo{title}{A theory of glacier dynamics and instabilities part
		1: Topographically confined glaciers}.
	\newblock \emph{\bibinfo{journal}{Journal of Glaciology}}
	\textbf{\bibinfo{volume}{68}}, \bibinfo{pages}{1--12} (\bibinfo{year}{2022}).
	
	\bibitem{benn2019}
	\bibinfo{author}{Benn, D.~I.}, \bibinfo{author}{Fowler, A.~C.},
	\bibinfo{author}{Hewitt, I.} \& \bibinfo{author}{Sevestre, H.}
	\newblock \bibinfo{title}{A general theory of glacier surges}.
	\newblock \emph{\bibinfo{journal}{Journal of Glaciology}}
	\textbf{\bibinfo{volume}{65}}, \bibinfo{pages}{701--716}
	(\bibinfo{year}{2019}).
	\newblock \urlprefix\url{https://doi.org/10.1017/jog.2019.62}.
	
	\bibitem{benn2023}
	\bibinfo{author}{Benn, D.~I.}, \bibinfo{author}{Hewitt, I.~J.} \&
	\bibinfo{author}{Luckman, A.~J.}
	\newblock \bibinfo{title}{Enthalpy balance theory unifies diverse glacier surge
		behaviour}.
	\newblock \emph{\bibinfo{journal}{Annals of Glaciology}}
	\textbf{\bibinfo{volume}{63}}, \bibinfo{pages}{88--94}
	(\bibinfo{year}{2023}).
	\newblock \urlprefix\url{https://doi.org/10.1017/aog.2023.23}.
	
	\bibitem{thogersen2019}
	\bibinfo{author}{Th{\o}gersen, K.}, \bibinfo{author}{Gilbert, A.},
	\bibinfo{author}{Schuler, T.~V.} \& \bibinfo{author}{Malthe-S{\o}renssen, A.}
	\newblock \bibinfo{title}{Rate-and-state friction explains glacier surge
		propagation}.
	\newblock \emph{\bibinfo{journal}{Nature communications}}
	\textbf{\bibinfo{volume}{10}}, \bibinfo{pages}{2823} (\bibinfo{year}{2019}).
	\newblock \urlprefix\url{https://doi.org/10.1038/s41467-019-10506-4}.
	
	\bibitem{gardner2018}
	\bibinfo{author}{Gardner, A.~S.} \emph{et~al.}
	\newblock \bibinfo{title}{Increased west antarctic and unchanged east antarctic
		ice discharge over the last 7 years}.
	\newblock \emph{\bibinfo{journal}{The Cryosphere}}
	\textbf{\bibinfo{volume}{12}}, \bibinfo{pages}{521--547}
	(\bibinfo{year}{2018}).
	\newblock \urlprefix\url{https://tc.copernicus.org/articles/12/521/2018/}.
	
	\bibitem{gardner2025}
	\bibinfo{author}{Gardner, A.~S.} \emph{et~al.}
	\newblock \bibinfo{title}{{ITS\_LIVE global glacier velocity data in near real
			time}}.
	\newblock \emph{\bibinfo{journal}{EGUsphere}} \textbf{\bibinfo{volume}{2025}},
	\bibinfo{pages}{1--29} (\bibinfo{year}{2025}).
	\newblock
	\urlprefix\url{https://egusphere.copernicus.org/preprints/2025/egusphere-2025-392/}.
	
	\bibitem{Shaman1988}
	\bibinfo{author}{Shaman, P.} \& \bibinfo{author}{Stine, R.~A.}
	\newblock \bibinfo{title}{The bias of autoregressive coefficient estimators}.
	\newblock \emph{\bibinfo{journal}{Journal of the American Statistical
			Association}} \textbf{\bibinfo{volume}{83}}, \bibinfo{pages}{842--848}
	(\bibinfo{year}{1988}).
	
	\bibitem{smith2026_code}
	\bibinfo{author}{Smith, T.}
	\newblock \bibinfo{title}{Estimating the resilience of non-stationary systems}
	(\bibinfo{year}{2026}).
	\newblock \urlprefix\url{https://doi.org/10.5281/zenodo.19731234}.
	
	\bibitem{Frisch1933}
	\bibinfo{author}{Frisch, R.} \& \bibinfo{author}{Waugh, F.~V.}
	\newblock \bibinfo{title}{Partial time regressions as compared with individual
		trends}.
	\newblock \emph{\bibinfo{journal}{Econometrica: Journal of the Econometric
			Society}} \bibinfo{pages}{387--401} (\bibinfo{year}{1933}).
	\newblock \urlprefix\url{https://doi.org/10.2307/1907330}.
	
	\bibitem{modlc}
	\bibinfo{author}{Friedl, M.} \& \bibinfo{author}{Sulla-Menashe, D.}
	\newblock \bibinfo{title}{Modis/terra+aqua land cover type yearly l3 global
		500m sin grid v061 [data set]}.
	\newblock \emph{\bibinfo{journal}{NASA EOSDIS Land Processes DAAC, accessed Jan
			2023}}  (\bibinfo{year}{2022}).
	\newblock \urlprefix\url{https://doi.org/10.5067/MODIS/MCD12Q1.061}.
	
	\bibitem{rgi}
	\bibinfo{author}{{RGI 7.0 Consortium}}.
	\newblock \bibinfo{title}{Randolph glacier inventory - a dataset of global
		glacier outlines, version 7.0}.
	\newblock \emph{\bibinfo{journal}{NSIDC: National Snow and Ice Data Center}}
	(\bibinfo{year}{2023}).
	\newblock \urlprefix\url{https://doi.org/10.5067/f6jmovy5navz}.
	
	\bibitem{ticoi}
	\bibinfo{author}{Charrier, L.} \emph{et~al.}
	\newblock \bibinfo{title}{Ticoi: an operational python package to generate
		regular glacier velocity time series}.
	\newblock \emph{\bibinfo{journal}{The Cryosphere}}
	\textbf{\bibinfo{volume}{19}}, \bibinfo{pages}{4555--4583}
	(\bibinfo{year}{2025}).
	
	\bibitem{gorelick2017}
	\bibinfo{author}{Gorelick, N.} \emph{et~al.}
	\newblock \bibinfo{title}{Google earth engine: Planetary-scale geospatial
		analysis for everyone}.
	\newblock \emph{\bibinfo{journal}{Remote Sensing of Environment}}
	\textbf{\bibinfo{volume}{202}}, \bibinfo{pages}{18--27}
	(\bibinfo{year}{2017}).
	\newblock \urlprefix\url{https://doi.org/10.1016/j.rse.2017.06.031}.
	\newblock \bibinfo{note}{Big Remotely Sensed Data: tools, applications and
		experiences}.
	
\end{thebibliography}
\end{document}